\documentclass[aps,prb,twocolumn,10pt,superscriptaddress,floatfix]{revtex4-1}

\usepackage[utf8]{inputenc}
\usepackage{amsmath,amsfonts,amssymb}
\usepackage{graphicx}
\usepackage{multirow}
\usepackage{array}
\usepackage{bm}
\usepackage{xcolor}
\usepackage{booktabs}

\begin{document}

\title{Ni ordering in the van der Waals gap of Ni-intercalated 2H-NbS$_2$: from disorder through $2\times2$ and $\sqrt{3}\times\sqrt{3}$ order to stripe order}

\author{Naveen Singh Dhami}
\email{naveen.dhami@universite-paris-saclay.fr}
\affiliation{Universit\'e Paris-Saclay, CNRS, Laboratoire de Physique des Solides, 91405 Orsay, France}
\affiliation{Synchrotron SOLEIL, L'Orme des Merisiers, Saint Aubin BP 48, 91192 Gif-sur-Yvette, France}

\author{Ricardo Guillermo Torres Ram\'irez}
\affiliation{Synchrotron SOLEIL, L'Orme des Merisiers, Saint Aubin BP 48, 91192 Gif-sur-Yvette, France}

\author{Gaurav Pransu}
\affiliation{Institute of Physics, Bijeni\v{c}ka cesta 46, 10000 Zagreb, Croatia}

\author{Ivo Batisti\'c}
\affiliation{Department of Physics, Faculty of Science, University of Zagreb, Bijeni\v{c}ka 32, 10000 Zagreb, Croatia}

\author{Pierre Fertey}
\affiliation{Synchrotron SOLEIL, L'Orme des Merisiers, Saint Aubin BP 48, 91192 Gif-sur-Yvette, France}

\author{Erik Elkaim}
\affiliation{Synchrotron SOLEIL, L'Orme des Merisiers, Saint Aubin BP 48, 91192 Gif-sur-Yvette, France}

\author{El-Eulmi Bendeif}
\affiliation{Synchrotron SOLEIL, L'Orme des Merisiers, Saint Aubin BP 48, 91192 Gif-sur-Yvette, France}

\author{Bruno Gudac}
\affiliation{Institute of Physics, Bijeni\v{c}ka cesta 46, 10000 Zagreb, Croatia}

\author{Nikolina Penić}
\affiliation{Department of Physics, Faculty of Science, University of Zagreb, Bijeni\v{c}ka 32, 10000 Zagreb, Croatia}

\author{Naveen Kumar Chogondahalli Muniraju
}
\affiliation{The Henryk Niewodnicza\'nski Institute of Nuclear Physics, Polish Academy of Sciences, ul. Radzikowskiego 152, 31-342 Krak\'ow, Poland}

\author{Neven Bari\v{s}i\'c}
\affiliation{Department of Physics, Faculty of Science, University of Zagreb, Bijeni\v{c}ka 32, 10000 Zagreb, Croatia}
\affiliation{Institute of Solid State Physics, TU Wien, 1040 Vienna, Austria}

\author{Petar Pop\v{c}evi\'c}
\email{ppopcevic@ifs.hr}
\affiliation{Institute of Physics, Bijeni\v{c}ka cesta 46, 10000 Zagreb, Croatia}

\author{Victor Bal\'edent}
\email{victor.baledent@universite-paris-saclay.fr}
\affiliation{Universit\'e Paris-Saclay, CNRS, Laboratoire de Physique des Solides, 91405 Orsay, France}
\affiliation{Institut Universitaire de France (IUF), Paris, France}

\begin{abstract}
We report a synchrotron X-ray diffraction study of Ni-intercalated 2H-NbS$_2$ over the range $0.18 \le x \le 0.57$, with Ni contents measured by energy-dispersive X-ray spectroscopy. Powder diffraction on the whole series shows that the NbS$_2$ host is preserved, while the Ni ions in the van der Waals gap evolve from a disordered arrangement to $2\times2$ order at $x=1/4$ and $\sqrt3\times\sqrt3$ order at $x=1/3$, through a largely disordered range above $x=1/3$, in which the excess Ni occupies the vacant sites of the $\sqrt3\times\sqrt3$ order, to an orthorhombic stripe order at $x=0.57$. The in-plane lattice parameter increases with Ni content, whereas the interlayer spacing remains nearly constant except for a pronounced minimum at $x=1/4$, which density-functional calculations reproduce for the ordered structure. Single-crystal diffraction on the two commensurate phases shows that the $x=1/4$ structure, with Ni on a single site and a trimerized Nb lattice, is free of stacking faults, whereas the chiral $x=1/3$ structure contains Ni on the nominally empty site, inversion twins and stacking faults of the Ni layers. The comparison between powders and crystals shows that grinding mainly degrades the stacking of the Ni layers. These results establish the structural phase diagram of Ni$_x$NbS$_2$ and the intercalant disorder to be taken into account in the interpretation of its magnetic properties, including the altermagnetic states predicted for the ordered phases.
\end{abstract}

\maketitle

\section{Introduction}

Transition metal dichalcogenides (TMDs) of general formula $TX_2$, where $T$ is a transition metal and $X$ a chalcogen, form one of the most versatile families of layered van der Waals materials. Within each layer, $T$ and $X$ are strongly covalently bonded, while adjacent $TX_2$ layers are held together only by weak van der Waals forces~\cite{han2018}. This anisotropy makes TMDs easy to exfoliate down to the monolayer, but it also opens a second, less exploited route to tailoring their properties: the intercalation of foreign atoms into the van der Waals gap. Because the layers are only weakly coupled, intercalation can proceed over a wide and continuously tunable range of concentrations without destroying the $TX_2$ sublattice, in contrast to substitution within the layers themselves~\cite{friend1987,wang_smartmat}.

Among the group-V TMDs, 2H-NbS$_2$ is a structurally simple metallic host: it crystallizes in the hexagonal space group $P6_3/mmc$ (No.~194), remains metallic down to low temperature without the charge-density-wave instability found in several isoelectronic compounds~\cite{naito1982}, and becomes superconducting below $T_c \approx 6$~K~\cite{vanmaaren1966,tissen2013}. This makes it a favorable platform to study how intercalated $3d$ ions order in the van der Waals gap, and how this ordering feeds back onto the host lattice.

Intercalation of $3d$ transition metals ($T' =$ V, Cr, Mn, Fe, Co, Ni) into NbS$_2$ and the related NbSe$_2$, TaS$_2$ and TaSe$_2$ hosts is known to stabilize commensurate superstructures in which the intercalant occupies octahedral sites of the van der Waals gap in an ordered fashion, most commonly at $x=1/3$ ($\sqrt{3}\times\sqrt{3}\,R30^\circ$, chiral space group $P6_322$, $AB$ stacking of the intercalant layers) and $x=1/4$ ($2\times2$, centrosymmetric space group $P6_3/mmc$, $AA$ stacking)~\cite{friend1987,vdB68,Anz70,morosan2007,horibe2014,craig2025}. These compounds host a remarkably diverse set of magnetic ground states, from N\'eel and stripe antiferromagnetism to chiral and frustrated triangular-lattice magnetism, strongly coupled to the metallic host bands~\cite{popcevic2022,an2023,boucher2024,xie2022}. Most studies, however, fix the intercalation level at a single commensurate value and do not address how the intercalant order itself develops as $x$ is varied. Composition series have been reported for Co$_x$NbS$_2$~\cite{nakayama2006} and, more recently, for Fe$_x$NbSe$_2$~\cite{xu2026}, but in the former the Co content was estimated from the nominal synthesis ratio rather than measured. For Ni$_x$NbS$_2$, a recent report further suggests that the host itself may depart from the pure 2H polytype at low intercalation levels, with a mixed 2H/3R stacking for $x \lesssim 0.15$~\cite{ramakrishna2025}.

In this work we present a systematic structural study of Ni$_x$NbS$_2$ powders and single crystals whose compositions, $0.18 \le x \le 0.57$, were determined independently by energy-dispersive X-ray spectroscopy (EDX). We first use high-resolution synchrotron powder diffraction on the whole series (Sec.~\ref{sec:powderall}) to identify the compositions at which $2\times2$ and $\sqrt{3}\times\sqrt{3}$ Ni superstructures appear, coexist, or give way to disorder and to a stripe phase, to follow the lattice parameters, compared with density-functional calculations of the ordered structures, and to estimate the correlation lengths of the Ni order. Single-crystal diffraction on the two commensurate phases $x=1/4$ and $x=1/3$ (Sec.~\ref{sec:SCall}) then provides their full structures, the widths of their superstructure reflections and their diffuse scattering. We finally discuss what the single crystals add to the powder picture and where the two differ (Sec.~\ref{sec:discussion}). Together, these results establish the structural phase diagram of Ni$_x$NbS$_2$ as the basis for a composition-dependent study of its transport and magnetic properties. These structural details are also directly relevant to altermagnetism, which has recently been reported in intercalated TMDs such as Co$_{1/4}$NbSe$_2$ and Co$_{1/4}$TaSe$_2$~\cite{regmi2025altermagnetism,dale2024,devita2025,sprague2026}. In an altermagnet, the spin splitting of the electronic bands arises neither from a net magnetization nor from spin--orbit coupling, but from a crystal symmetry that relates the magnetic sublattices of opposite spin by a rotation or a mirror rather than by a translation or an inversion~\cite{smejkal2022a,smejkal2022b}. Whether such a state can exist therefore depends on the order, the stacking and the chirality of the magnetic intercalant layers, which are the subject of this work (Sec.~\ref{sec:altermag}).

\section{Methods}
\label{sec:methods}

\subsection{Crystal growth and composition}
Single crystals of Ni$_x$NbS$_2$ were grown by chemical vapor transport (CVT) from the elements in the desired stoichiometric ratio, sealed in evacuated quartz ampoules with iodine as transport agent. A temperature gradient of 50~K was applied along the ampoule, the hot end being held at 950$^\circ$C~\cite{ubaldini2014}. One batch was grown per composition, except for $x=1/4$ and $x=1/3$, for which several batches were grown and the best crystals are reported here. The crystals have a well-defined hexagonal habit with shiny black $ab$ faces, lateral dimensions from a few micrometers to a centimeter, and thicknesses from a few micrometers to about half a millimeter; they cleave readily perpendicular to $c$.

The composition and morphology of the crystals were examined by scanning electron microscopy (SEM) using a Zeiss GeminiSEM~300 equipped with an Everhart--Thornley secondary-electron detector and an Oxford Instruments energy-dispersive X-ray spectroscopy (EDX) detector. Flat crystals were mounted on aluminum stubs with carbon tape, and EDX maps (typically $100\times100~\mu$m$^2$) were recorded at 20~kV at several positions on several crystals of each batch; quantification used the Oxford Instruments AZtec software. These measurements yielded the compositions $x = 0.18$, 0.25, 0.33, 0.37, 0.46 and 0.57 used throughout this work.

\subsection{Powder X-ray diffraction}
Powder X-ray diffraction was measured at room temperature on the two-circle diffractometer of the CRISTAL beamline at Synchrotron SOLEIL, equipped with Mythen2 silicon-strip detectors (Dectris), at a wavelength $\lambda = 0.6779$~\AA\ (18.29~keV) calibrated against a LaB$_6$ standard, over $Q = 0.1$--$8.4$~\AA$^{-1}$. The powders were obtained by grinding crystals from the same growth batches as the single crystals. Lattice parameters, Ni occupancies and phase fractions were obtained by Rietveld refinement with Profex/BGMN~\cite{doebelin2015} (examples are shown in Fig.~S5 of the SM~\cite{SM}), using the parent $P6_3/mmc$ cell and, where the corresponding superlattice reflections are observed, the $2\times2$ ($P6_3/mmc$) and $\sqrt{3}\times\sqrt{3}$ ($P6_322$) supercells. In these refinements the total Ni occupancy was refined, while the distribution of Ni over the octahedral sites was imposed by the structural model: the powder data determine the Ni content but not the degree of Ni order. All lattice parameters are quoted in the parent hexagonal cell.

\subsection{Single-crystal X-ray diffraction}
Single crystals of about 30~$\mu$m, preselected at the Laboratoire de Physique des Solides for their low mosaicity, were measured at $T = 294$~K on the four-circle Newport diffractometer of CRISTAL equipped with a Rigaku Oxford Diffraction Atlas CCD detector, at $\lambda = 0.6433$~\AA. A total of 9616 and 10361 reflections were collected for $x=1/4$ and $x=1/3$, respectively, reducing to 229 and 468 independent reflections ($R_{\rm int} =$ $0.075$ and $0.042$) up to $\theta_{\max} \simeq$ 24.6$^\circ$ ($d_{\min} \simeq$ 0.77~\AA) for $x=1/4$ and $\theta_{\max} \simeq 30.7^\circ$ ($d_{\min} \simeq 0.63$~\AA) for $x=1/3$. The data were integrated and scaled with CrysAlis$^{\rm Pro}$ (Rigaku Oxford Diffraction), including an empirical spherical-harmonics absorption correction (SCALE3 ABSPACK) . The structures were solved with SHELXT~\cite{sheldrick2015a} and refined by full-matrix least squares on $F^2$ with SHELXL~\cite{sheldrick2015} within Olex2~\cite{dolomanov2009}. For $x=1/4$ all site occupancies except that of Ni were fixed at unity, with Ni on the $2a$ site only (19 refined parameters: scale, Ni occupancy, positions and anisotropic displacement parameters). For $x=1/3$ the occupancies of the two Ni sites, $2d$ and $2b$, were refined with their sum constrained to $x=1/3$ (19 parameters), all other occupancies being fixed at unity. Reciprocal-space planes were reconstructed from the same data sets with CrysAlis$^{\rm Pro}$.

Peak widths were extracted from one-dimensional cuts through the reconstructed planes, integrated over $\pm0.1$~r.l.u.\ in the perpendicular direction and fitted with a Gaussian plus a constant background. All host and superstructure reflections with a signal-to-noise ratio above 10 were analyzed (141 host and 61 superstructure reflections for $x=1/4$, 55 and 181 for $x=1/3$). The full widths at half maximum (FWHM) were converted into $\Delta q$ using $\Delta q_L = (2\pi/c)\,\mathrm{FWHM}_L$ along $c^*$, and $\Delta q_H = (4\pi/\sqrt3 a)\,\mathrm{FWHM}_H$ for cuts along $(H,0,0)$ ($x=1/4$) or $\Delta q_H = (4\pi/a)\,\mathrm{FWHM}_H$ for cuts along $(H,H,0)$ ($x=1/3$).

\subsection{Density-functional calculations}
Density-functional theory (DFT) calculations were performed with Quantum ESPRESSO~\cite{giannozzi2009,giannozzi2017} using the PBEsol exchange--correlation functional~\cite{perdew2008}, ultrasoft pseudopotentials from pslibrary~1.0.0~\cite{dalcorso2014}, Marzari--Vanderbilt smearing of 0.01~Ry~\cite{marzari1999} and the simplified rotationally invariant DFT+$U$ scheme~\cite{dudarev1998}. A Hubbard parameter $U = 7.0$~eV was applied to the Ni $3d$ states for all compositions, close to the value $U = 7.7$~eV obtained for $x = 1/3$ from linear-response density-functional perturbation theory~\cite{timrov2018,timrov2021,timrov2022}. Plane-wave cutoffs of 100--110~Ry for the wave functions and 900~Ry for the charge density were used, with $k$-point meshes of $5\times5\times5$ ($3\times3$ supercell, $x=1/9$), $11\times11\times6$ ($x=1/4$) and $12\times12\times6$ ($x=1/3$); the stripe order at $x=1/2$ was modeled in the orthohexagonal $a\times\sqrt3a\times c$ cell, with Ni rows along $a$ stacked $AA$ along $c$ . Lattice parameters and atomic positions of the ordered supercells were fully relaxed.

\section{Powder diffraction across the series}
\label{sec:powderall}

\subsection{Composition and Ni superstructures}
\label{sec:composition}
\label{sec:powder}
\label{sec:xi}

\begin{figure}[tbp]
	\centering
	\includegraphics[width=0.85\linewidth]{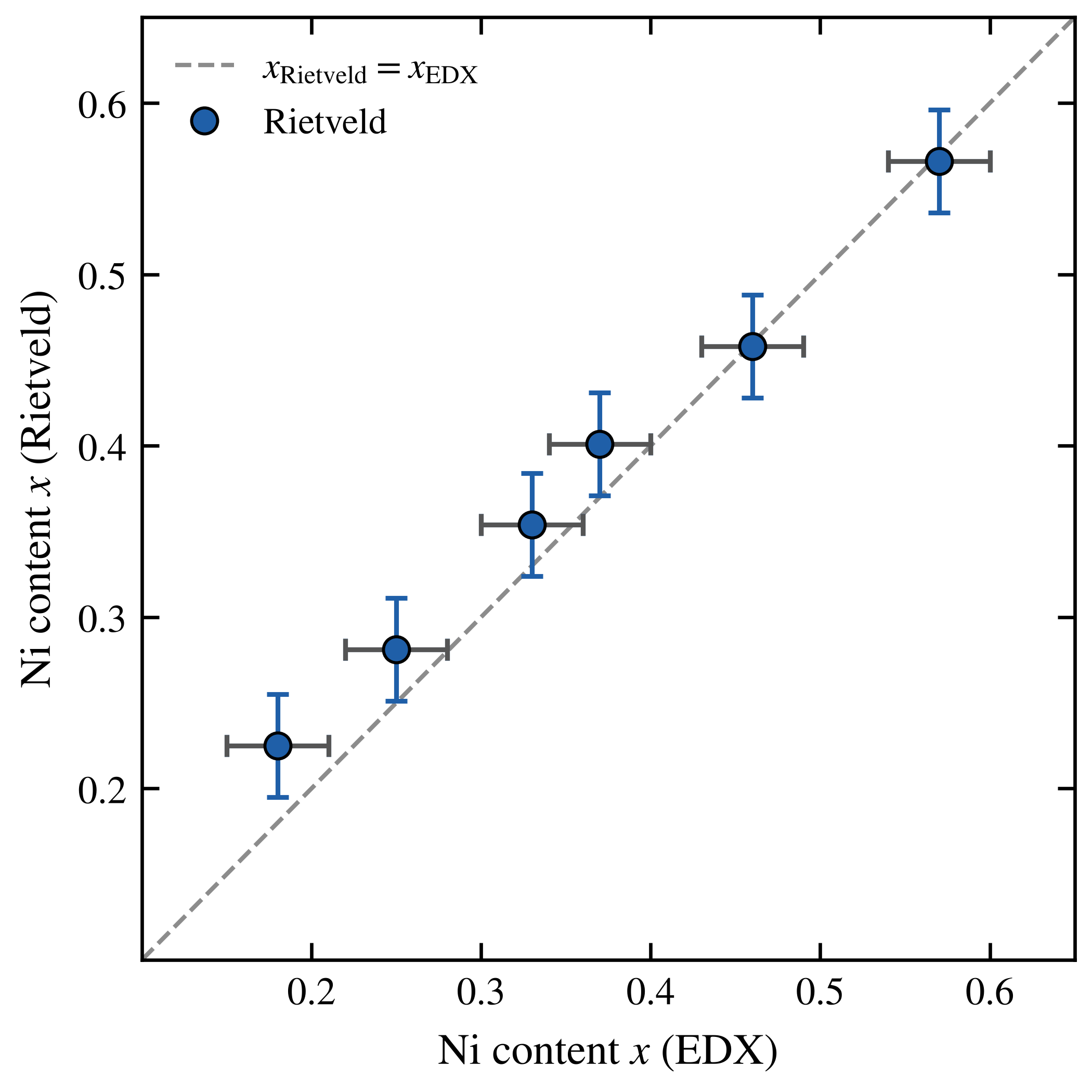}
    \caption{Ni content $x$ of Ni$_x$NbS$_2$ obtained from Rietveld refinement of the synchrotron powder diffraction data versus $x$ determined by EDX. The dashed line indicates equality. Error bars ($\pm0.03$) reflect the correlation between Ni occupancy and displacement parameters in the refinement.}
    \label{Fig:EDX}
\end{figure}

Figure~\ref{Fig:EDX} compares the Ni content obtained by EDX with the total Ni occupancy refined from the powder diffraction data. The two determinations agree within the uncertainty of the refinement ($\pm0.03$, dominated by the correlation between Ni occupancy and displacement parameters) for $x \ge 0.25$, the refined value being systematically slightly larger than the EDX one at low $x$; the largest deviation, $0.045$, occurs for $x=0.18$. In the following, compositions are labelled by their EDX values.

\begin{figure}[tbp]
	\centering
	\includegraphics[width=1.0\linewidth]{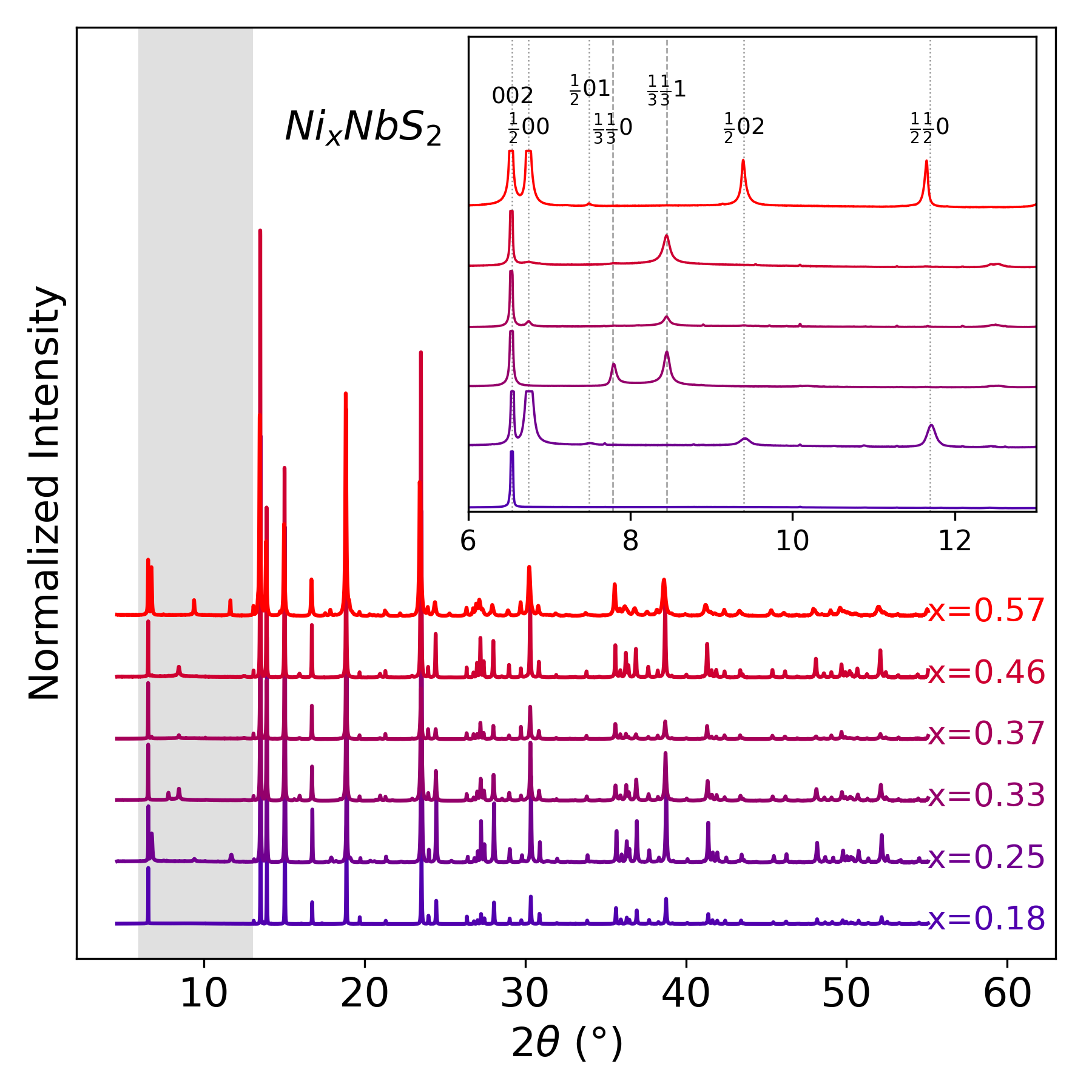}
    \caption{Synchrotron powder diffraction patterns of Ni$_x$NbS$_2$ at room temperature ($\lambda = 0.6779$~\AA), from $x=0.18$ (bottom) to $x=0.57$ (top). A smooth background was subtracted from each pattern, which was then normalized to the intensity of the $002$ reflection; the vertical offset is arbitrary. The shaded band indicates the range magnified in the inset. Inset: $6$--$13^\circ$ range with the intensity multiplied by three, the strong $002$ and $\frac{1}{2}00$ reflections being truncated. Vertical lines mark the calculated positions of the host $002$ reflection and of the superlattice reflections, indexed in the parent hexagonal cell: dotted lines for the $2\times2$ superstructure (half-integer indices) and dashed lines for the $\sqrt3\times\sqrt3\,R30^\circ$ superstructure (indices in thirds), computed with the mean cell of the series ($a = 3.327$~\AA, $c = 11.888$~\AA); the corresponding reflections shift by at most $0.07^\circ$ across the series.}
    \label{Fig:PXRD}
\end{figure}

The diffraction patterns of the whole series are shown in Fig.~\ref{Fig:PXRD}. All of them are indexed by the 2H-NbS$_2$ host cell, supplemented by superlattice reflections arising from the arrangement of Ni in the van der Waals gaps. Although such superstructures are usually resolved by single-crystal diffraction, the brilliance and angular resolution of CRISTAL allow the weak superlattice reflections to be identified unambiguously in the powder patterns, so that both the $2\times2$ (half-integer indices) and the $\sqrt{3}\times\sqrt{3}\,R30^\circ$ (indices in thirds) motifs can be followed across the series (inset of Fig.~\ref{Fig:PXRD} and Table~\ref{Tab:lattice}). A small 3R-NbS$_2$ fraction is detected at $x=0.18$ only, and traces of elemental sulfur ($\alpha$-S$_8$, $\lesssim0.1$~wt\%) in most samples.

At $x=0.18$, no superlattice reflection is observed, consistent with a disordered Ni occupancy of the van der Waals gap. At $x=0.25$, only the half-integer reflections of the $2\times2$ order are present, and at $x=0.33$ only the reflections in thirds of the $\sqrt{3}\times\sqrt{3}$ order, with both even and odd $\ell$ ($\tfrac13\tfrac13 0$ and $\tfrac13\tfrac13 1$ in the inset of Fig.~\ref{Fig:PXRD}). Above $x=1/3$, however, the superlattice reflections no longer correspond to a fully ordered structure. Normalized to the integrated intensity of a host reflection of similar orientation ($100$ for the $hk0$ reflections, $102$ for the $hk1$ ones), which limits the effect of preferred orientation, the integrated intensity of a given superlattice reflection can be compared from one sample to the next (Table~S4 of the SM~\cite{SM}). At $x=0.46$, the $\tfrac13\tfrac13 1$ reflection is even slightly stronger than at $x=0.33$, by 10\%, whereas $\tfrac13\tfrac13 0$ drops to 2\% of its intensity at $x=0.33$, with an unchanged width. At $x=0.37$, $\tfrac13\tfrac13 1$ is reduced to 60\% of its intensity at $x=0.33$, and $\tfrac13\tfrac13 0$ again to 2\%. Within each pattern, independently of any normalization, the ratio $I(\tfrac13\tfrac13 0)/I(\tfrac13\tfrac13 1)$ falls from 0.51 at $x=0.33$ to 0.022 at $x=0.37$ and 0.015 at $x=0.46$. Both samples also show weak $2\times2$ reflections, the strongest being $\tfrac12 00$, at 12\% ($x=0.37$) and 5\% ($x=0.46$) of its intensity at $x=0.25$. Above $x=1/3$, the $\sqrt3\times\sqrt3$ order is therefore seen almost only through its odd-$\ell$ reflections, and at $x=0.37$ neither order is well developed: in this range, a large part of the Ni is not long-range ordered. The selective loss of the even-$\ell$ reflections shows where the excess Ni goes, as discussed in Sec.~\ref{sec:distribution}. The strong preferred orientation inherent to the platelet morphology prevented a reliable determination of the volume fractions of the two orders in this range; it also affects the intensity ratio of reflections with different orientations, such as $hk0$ and $hk1$, within a pattern, but much less the comparison of the same reflection between samples. At $x=0.57$, finally, strong half-integer reflections reappear, narrower than at $x=0.25$ and with a different distribution of intensities, while the host reflections with in-plane components split into doublets (Sec.~\ref{sec:disc057}).

The superlattice reflections are also broader than the host reflections throughout the series. A multi-reflection analysis of the excess widths [Supplemental Material (SM)~\cite{SM}] gives in-plane correlation lengths of the Ni order of several hundred \AA\ at all ordered compositions, but correlation lengths along $c$ of only $\approx100$--$200$~\AA, i.e.\ about ten unit cells, for $0.25 \le x \le 0.46$, while the host is correlated over $\approx500$--$650$~\AA\ in plane and $\approx1000$--$1500$~\AA\ along $c$ (at $x=0.37$ only two superlattice reflections are resolved, so that the in-plane and out-of-plane widths cannot be separated). The stacking of the Ni layers from one van der Waals gap to the next is therefore the least correlated element of the structure in the powders, and, as shown by the comparison with single crystals in Sec.~\ref{sec:SCvsP}, the one most affected by grinding. The exception is $x=0.57$, for which the superlattice reflections show no measurable excess broadening along $c$.

\subsection{Lattice parameters and comparison with DFT}
\label{sec:lattice}

\begin{table*}[t]
\caption{Lattice parameters of the host (parent hexagonal cell) and observed superstructures of the Ni$_x$NbS$_2$ powders, with $x$ determined by EDX. The statistical uncertainty of the lattice parameters is about $1\times10^{-4}$~\AA; the relative uncertainty between samples, used for the error bars of Fig.~\ref{Fig:LatticeParams}, is estimated at $1\times10^{-3}$~\AA, and the common wavelength-calibration uncertainty is 0.1--0.3\%. For $x=0.57$, $a$ is the hexagonal average of the orthorhombic cell.}
\label{Tab:lattice}
\begin{ruledtabular}
\begin{tabular}{ccccc}
$x$ & $a$ (\AA) & $c$ (\AA) & Superstructure & Impurities \\
\hline
0.18 & 3.3225 & 11.8850 & none & 3R-NbS$_2$, $\alpha$-S$_8$ \\
0.25 & 3.3226 & 11.8648 & $2\times2$ & $\alpha$-S$_8$ (weak), unidentified lines $\le 0.2\%$ \\
0.33 & 3.3255 & 11.8924 & $\sqrt{3}\times\sqrt{3}$ (even and odd $\ell$) & $\alpha$-S$_8$ (weakest) \\
0.37 & 3.3276 & 11.8939 & $\sqrt{3}\times\sqrt{3}$ (odd $\ell$ only) + weak $2\times2$ & $\alpha$-S$_8$ \\
0.46 & 3.3276 & 11.8940 & $\sqrt{3}\times\sqrt{3}$ (odd $\ell$ only) + very weak $2\times2$ & $\alpha$-S$_8$ \\
0.57 & 3.3344 & 11.8965 & $2\times2$, orthorhombic (stripes) & none detected \\
\end{tabular}
\end{ruledtabular}
\end{table*}

\begin{figure}[tbp]
	\centering
	\includegraphics[width=1.0\linewidth]{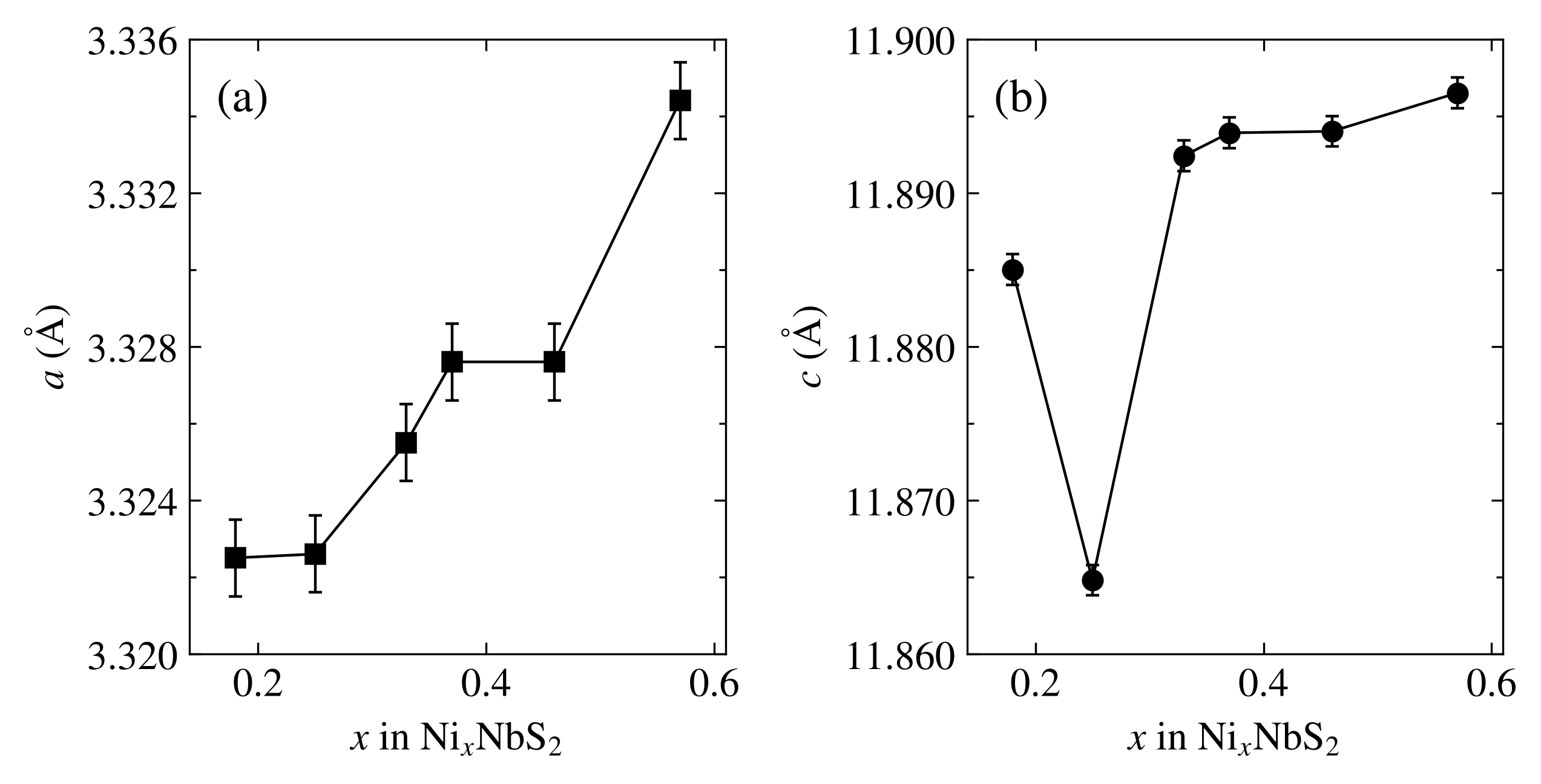}
	\caption{Lattice parameters of Ni$_x$NbS$_2$ in the parent hexagonal cell from Rietveld refinement of the powder data (Table~\ref{Tab:lattice}). (a) In-plane parameter $a$. (b) Out-of-plane parameter $c$, with a pronounced minimum at $x=0.25$.}
\label{Fig:LatticeParams}
\end{figure}

The refined lattice parameters are listed in Table~\ref{Tab:lattice} and shown in Fig.~\ref{Fig:LatticeParams}. The in-plane parameter $a$ increases with Ni content, from 3.3225~\AA\ at $x=0.18$ to 3.3344~\AA\ at $x=0.57$, reflecting a progressive expansion of the Nb--S framework, as reported for Co$_x$NbS$_2$~\cite{nakayama2006}; it is however constant between $x=0.18$ and $0.25$. The $c$ parameter varies by less than 0.035~\AA\ over the whole series. It is nearly constant from $x=0.33$ upward, at $11.894\pm0.003$~\AA, but shows a pronounced minimum at $x=0.25$, $0.020$~\AA\ below the value at $x=0.18$ and $0.028$~\AA\ below that at $x=0.33$. A similar but stronger $c$-axis response, saturating near $x=0.3$ with a slope about an order of magnitude larger, has been reported for Co$_x$NbS$_2$~\cite{nakayama2006}; a comparison with other intercalated TMDs is given in the SM~\cite{SM}. Whereas $c$ grows steadily upon Fe or Cu intercalation, the Ni and Co compounds of NbS$_2$ remain close to the pristine value.

Density-functional calculations of the relaxed ordered structures help to interpret these trends. Their lattice parameters are compared with experiment in Table~\ref{Tab:DFT}. As usual for PBEsol, $a$ and $c$ are slightly underestimated, by about 0.3--0.4\% and 0.1--0.45\%, respectively. The calculations reproduce the increase of $a$ with Ni content, up to the stripe structure at $x=1/2$. More remarkably, they also yield a local minimum of $c$ at $x=1/4$ for the perfectly ordered supercells: $c$ decreases by 0.008~\AA\ from $x=1/9$ to $x=1/4$ and increases by 0.003~\AA\ from $x=1/4$ to $x=1/3$. The sign of these variations is that of the experiment, but their amplitude is several times smaller than the measured $-0.020$ and $+0.028$~\AA\ (between $x=0.18$, 0.25 and 0.33). For the fully ordered stripe structure at $x=1/2$, the calculated $c$ is even shorter, 11.842~\AA: as at $x=1/4$, a perfectly ordered $2\times2$-derived Ni arrangement contracts the interlayer spacing. This contraction is not observed at $x=0.57$, whose $c$ lies on the plateau of the higher compositions; we come back to this difference in Sec.~\ref{sec:disc057}.

\begin{table}[tbp]
\centering
\caption{Lattice parameters (parent hexagonal cell, in \AA) of the relaxed ordered DFT+$U$ supercells compared with the powder values of Table~\ref{Tab:lattice} at the closest compositions.}
\label{Tab:DFT}
\setlength{\tabcolsep}{1pt}
\begin{ruledtabular}
\begin{tabular}{lcccc}
 & $x = 1/9$ & $x = 1/4$ & $x = 1/3$ & $x = 1/2$ \\
\hline
Cell (atoms)  & $3\times3$ (56) & $2\times2$ (26) & $\sqrt{3}\times\sqrt{3}$ (20) & $1\times\sqrt3$ (14) \\
$a$, DFT           & 3.309  & 3.313  & 3.316 & 3.322$^\dagger$ \\
$c$, DFT           & 11.858 & 11.850 & 11.853 & 11.842 \\
$a$, exp.          & 3.3225$^*$ & 3.3226 & 3.3255 & 3.3344$^\ddagger$ \\
$c$, exp.          & 11.8850$^*$ & 11.8648 & 11.8924 & 11.8965$^\ddagger$ \\
\end{tabular}
\end{ruledtabular}
{\footnotesize $^*$ $x=0.18$, disordered. $^\dagger$ Mean of $a_o = 3.338$~\AA\ and $b_o/\sqrt3 = 3.305$~\AA\ of the orthorhombic cell. $^\ddagger$ $x=0.57$; hexagonal average of the orthorhombic cell.}
\end{table}

The powder data thus establish the sequence of Ni orders across the series, the composition dependence of the lattice parameters with its minimum of $c$ at $x=0.25$, and the short correlation length of the Ni order along $c$. They leave three questions open. Since the distribution of Ni over the octahedral sites is imposed by the structural model in the Rietveld refinements, they tell how complete the Ni order is only indirectly, through the superlattice intensities. They cannot decide whether the broadening of the superstructure reflections is intrinsic or introduced by grinding. Finally, they give no access to the local structure, the Ni--S and Nb--Nb distances, needed to understand the minimum of $c$. These questions are addressed by single-crystal diffraction on the two commensurate phases.

\section{Single-crystal diffraction of the commensurate phases}
\label{sec:SCall}

\subsection{Refined structures}
\label{sec:SC}

\begin{figure}[tbp]
	\centering
	\includegraphics[width=0.9\linewidth]{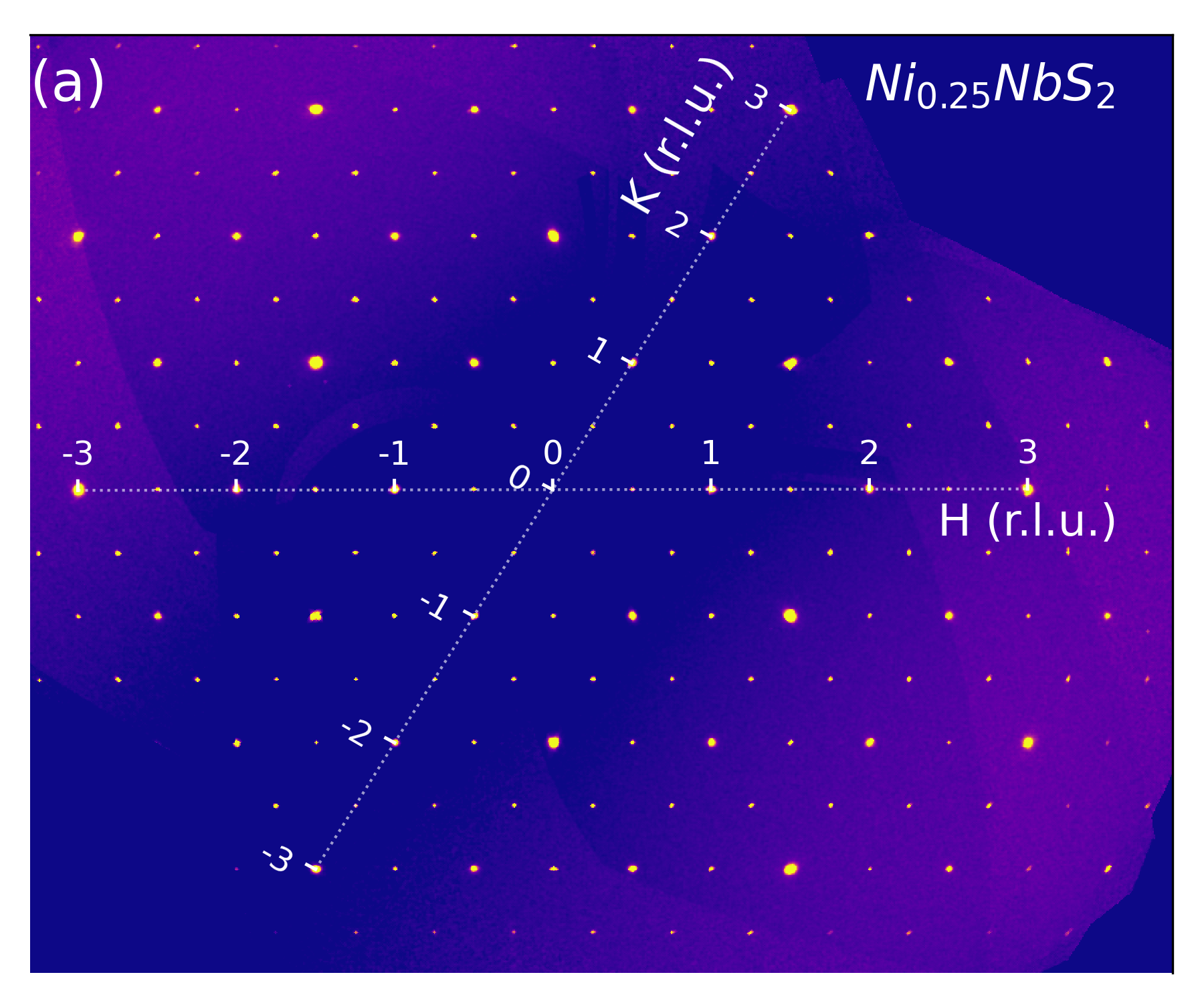}
	\includegraphics[width=0.9\linewidth]{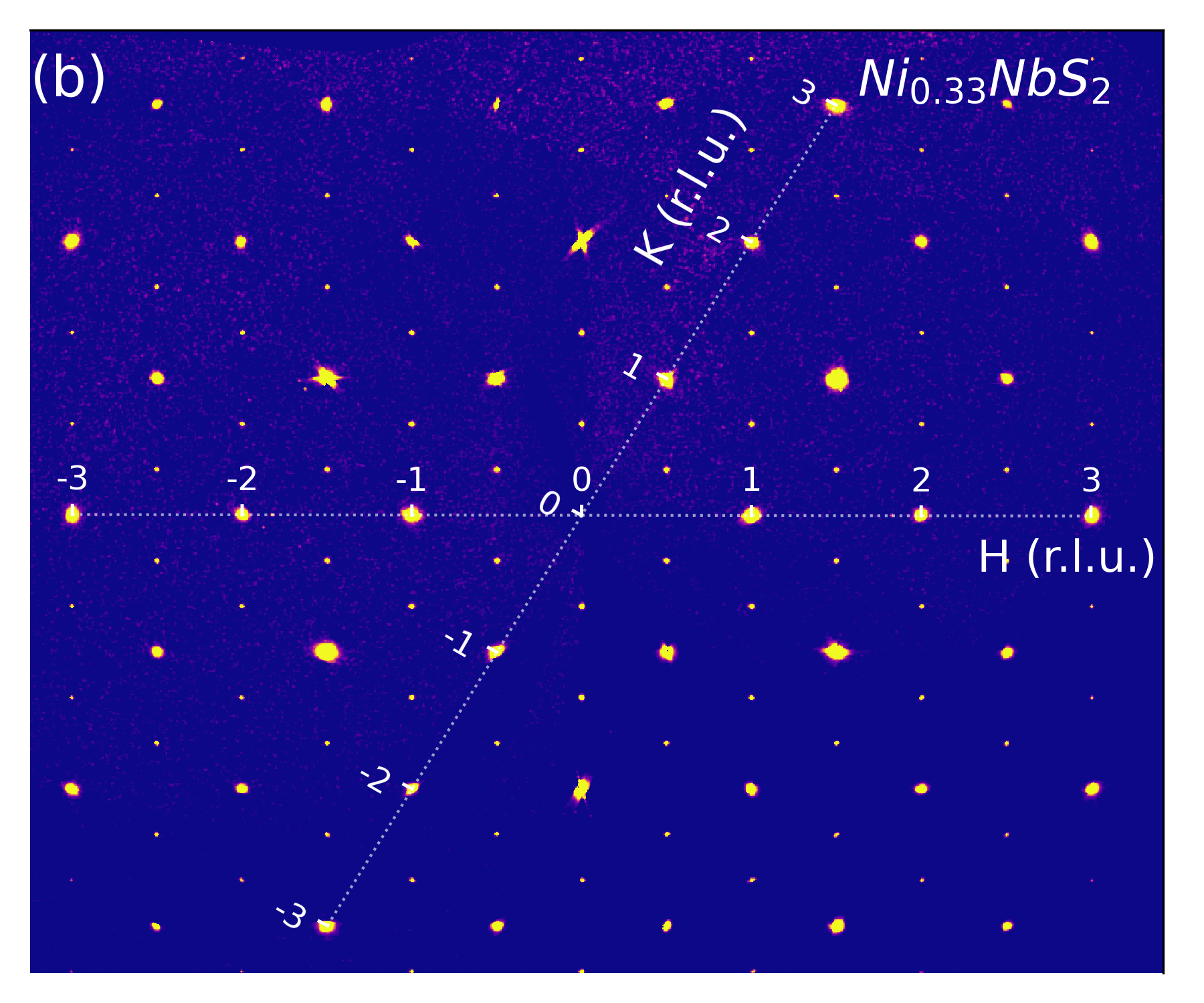}
	\caption{Reciprocal-space reconstructions of the $(H,K,0)$ plane of (a) Ni$_{1/4}$NbS$_2$ and (b) Ni$_{1/3}$NbS$_2$, indexed in the reciprocal lattice of the parent hexagonal cell ($a \simeq 3.31$~\AA). Superstructure reflections appear at half-integer $H,K$ for $x=1/4$ and at $H,K = n/3$ for $x=1/3$.}
\label{Fig:ReciprocalSpace}
\end{figure}

Given the singular behavior of the $x=0.25$ composition, single-crystal diffraction was performed on $x=1/4$ and on $x=1/3$, the two compositions for which the superstructure is best defined. The reciprocal-space reconstructions in the $(H,K,0)$ plane are shown in Fig.~\ref{Fig:ReciprocalSpace}, indexed in the reciprocal lattice of the parent hexagonal cell. Superstructure reflections appear at half-integer $H$ and $K$ for $x=1/4$, and at $H,K = n/3$ for $x=1/3$. The data are described by the space groups $P6_3/mmc$ (No.~194, $a' = 2a$) and $P6_322$ (No.~182, $a' = \sqrt3 a$), respectively. The refined structures are given in Tables~\ref{Tab:Refinement0p25} and~\ref{Tab:Refinement0p33}. They are isostructural with the corresponding members of the Fe$_x$TaS$_2$ family, in which the $2a\times2a$ order at $x=1/4$ is centrosymmetric with $AA$ stacking of the intercalant layers and the $\sqrt3a\times\sqrt3a$ order at $x=1/3$ is chiral with $AB$ stacking~\cite{morosan2007,horibe2014}.

For $x=1/4$, the structure was refined with Ni on the $2a$ octahedral site $(0,0,0)$ of the quadrupled cell only, the three other octahedral positions per van der Waals gap ($6g$) being left empty. All occupancies were fixed at unity except that of Ni, which refines to 0.944(8), i.e.\ $x=0.236(2)$, close to the EDX value and slightly below the total Ni content refined from the powder data, 0.28(3) (Fig.~\ref{Fig:EDX}). This single-site model describes the data with $R_1 = 2.56\%$, and the largest residual peak of the difference Fourier map, 0.99~e\,\AA$^{-3}$, lies close to the Nb1 column and not on the empty $6g$ sites; a small fraction of Ni on these sites, not included in the model, cannot be excluded. The absence of diffuse scattering at this composition (Sec.~\ref{sec:diffuse}) further shows that the stacking of the Ni layers is well correlated in this crystal. This order distorts the Nb sublattice: the Nb atoms split into a $6h$ site  and an undistorted $2b$ site lying in the same column as the Ni ions. The $6h$ Nb atoms form equilateral Nb$_3$ triangles with a short Nb--Nb distance of 3.2005(6)~\AA, against 3.4285~\AA\ between triangles and $a'/2 = 3.31449$~\AA\ for the undistorted lattice, i.e.\ a trimerization of the Nb triangular net of about $3.4\%$. The same analysis applied to the published structure of Fe$_{1/4}$TaS$_2$~\cite{morosan2007} gives short and long Ta--Ta distances of 3.209 and 3.405~\AA\ for $a'/2 = 3.307$~\AA, so that this trimerization is a generic response of the host layer to $2a\times2a$ intercalant order rather than a peculiarity of Ni$_{1/4}$NbS$_2$.

For $x=1/3$, in contrast, the Ni distribution is not confined to a single site. Besides the $2d$ site $(\tfrac13,\tfrac23,\tfrac34)$ of the $\sqrt3\times\sqrt3$ cell, which refines to an occupancy of 0.910(2), a minority Ni population of 0.090(2) is found on the $2b$ site (the sum of the two occupancies being constrained to the nominal content, so that only their partition is determined) $(0,0,\tfrac34)$, i.e.\ on the octahedral position that the ideal $\sqrt3\times\sqrt3$ order leaves vacant. About 9\% of the Ni therefore sits on the ``wrong'' site of the triangular Ni superlattice. The Ni--S distances of the two sites are close but not identical, 2.3796(4)~\AA\ for $2d$ and 2.3689(4)~\AA\ for $2b$, both corresponding to regular NiS$_6$ octahedra. The Nb sublattice remains essentially undistorted (Nb--Nb $=3.31796$ and 3.31767~\AA, against $a'/\sqrt3 = 3.31757$~\AA), in marked contrast with the trimerized Nb net at $x=1/4$. The Flack parameter refines to 0.39(7) [0.35(6) from Parsons' quotients~\cite{parsons2013}]: the measured crystal is not a single enantiomorph but contains an inversion-twin fraction of about one third.

\begin{table}[tbp]
  \centering
  \caption{Refined atomic positions of Ni$_{1/4}$NbS$_2$ at 294~K in space group $P6_3/mmc$ (No.~194) with $a = 6.62897(7)$~\AA, $c = 11.82484(16)$~\AA\ and $Z = 8$. Refinement on 229 independent reflections (226 with $I > 2\sigma(I)$, $R_{\rm int} = 0.075$) and 19 parameters converged to $R_1 = 2.56$\%, $wR_2 = 5.31$\%, GooF $= 1.35$. All occupancies were fixed at unity (not refined) except that of Ni; $U_{\rm eq}$ is the equivalent isotropic displacement parameter.}
  \label{Tab:Refinement0p25}
    \resizebox{\columnwidth}{!}{%
    \begin{tabular}{lc|ccccc}
      \hline
      Atom & Wyckoff & $x$ & $y$ & $z$ & occ. & $U_{\text{eq}}$ (\AA$^2$) \\
      \hline
      Ni  & $2a$  & 0           & 0           & 0        & 0.944(8) & 0.0094(3) \\
      Nb1 & $6h$  & 0.49427(3) & 0.98853(6) & 1/4        & 1 & 0.0064(3) \\
      Nb2 & $2b$  & 0           & 0           & 1/4        & 1 & 0.0065(3) \\
      S1  & $12k$ & 0.16764(4)   & 0.83236(4)   & 0.11897(7)  & 1 & 0.0077(3) \\
      S2  & $4f$  & 2/3         & 1/3         & 0.11436(8)  & 1 & 0.0078(4) \\
      \hline
    \end{tabular}}
\end{table}

\begin{table}[tbp]
  \centering
  \caption{Refined atomic positions of Ni$_{1/3}$NbS$_2$ at 294~K in the chiral space group $P6_322$ (No.~182) with $a = 5.74620(10)$~\AA, $c = 11.8657(2)$~\AA\ and $Z = 6$. Refinement on 468 independent reflections (465 with $I > 2\sigma(I)$, $R_{\rm int} = 0.042$) and 19 parameters converged to $R_1 = 2.06$\%, $wR_2 = 5.43$\%, GooF $= 1.324$, with a Flack parameter of 0.39(7). The Ni occupancies of the $2d$ and $2b$ sites were constrained to sum to $x = 1/3$, so that the refinement determines their partition and not the total Ni content.}
  \label{Tab:Refinement0p33}
    \resizebox{\columnwidth}{!}{%
    \begin{tabular}{lc|ccccc}
      \hline
       Atom & Wyckoff & $x$ & $y$ & $z$ & occ. & $U_{\text{eq}}$ (\AA$^2$) \\
      \hline
      Ni1 & $2d$  & 1/3         & 2/3         & 3/4         & 0.910(2) & 0.00740(15)  \\
      Ni2 & $2b$  & 0           & 0           &3/4         & 0.090(2) & 0.0057(12)$^*$ \\
      Nb1 & $4f$  & 2/3         & 1/3         & 0.50216(2)  & 1        & 0.00436(15)  \\
      Nb2 & $2a$  & 0           & 0           & 1/2         & 1        & 0.00431(15)  \\
      S   & $12i$ & 0.33276(7) & 0.33180(7) & 0.63183(4) & 1        & 0.00553(16)  \\
      \hline
    \end{tabular}}
    \\[2pt]
    {\footnotesize $^*$ refined isotropically ($U_{\rm iso}$).}
\end{table}

\subsection{Reflection widths and diffuse scattering}
\label{sec:widths}
\label{sec:diffuse}

\begin{figure}[tbp]
	\centering
	\includegraphics[width=0.99\linewidth]{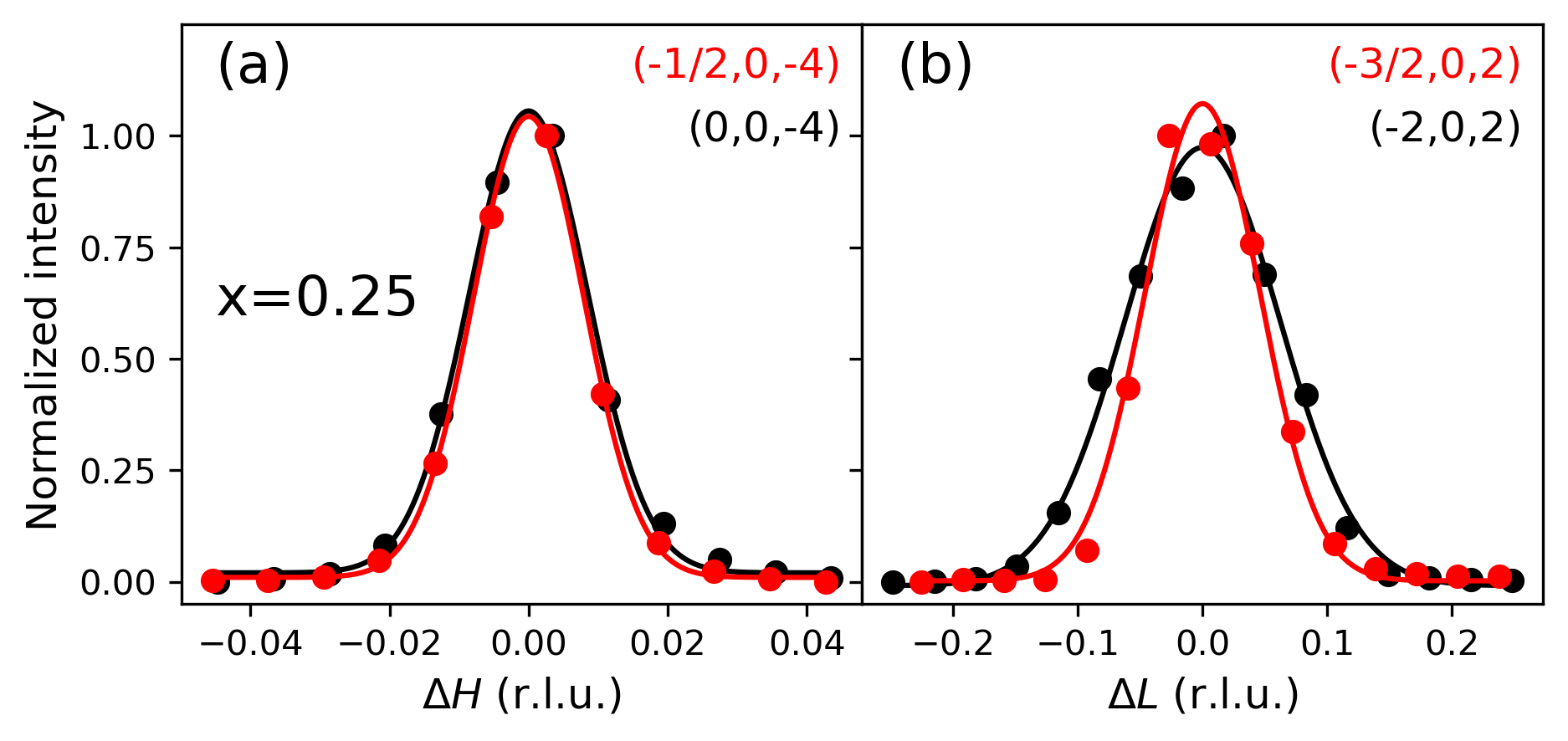}
	\includegraphics[width=0.99\linewidth]{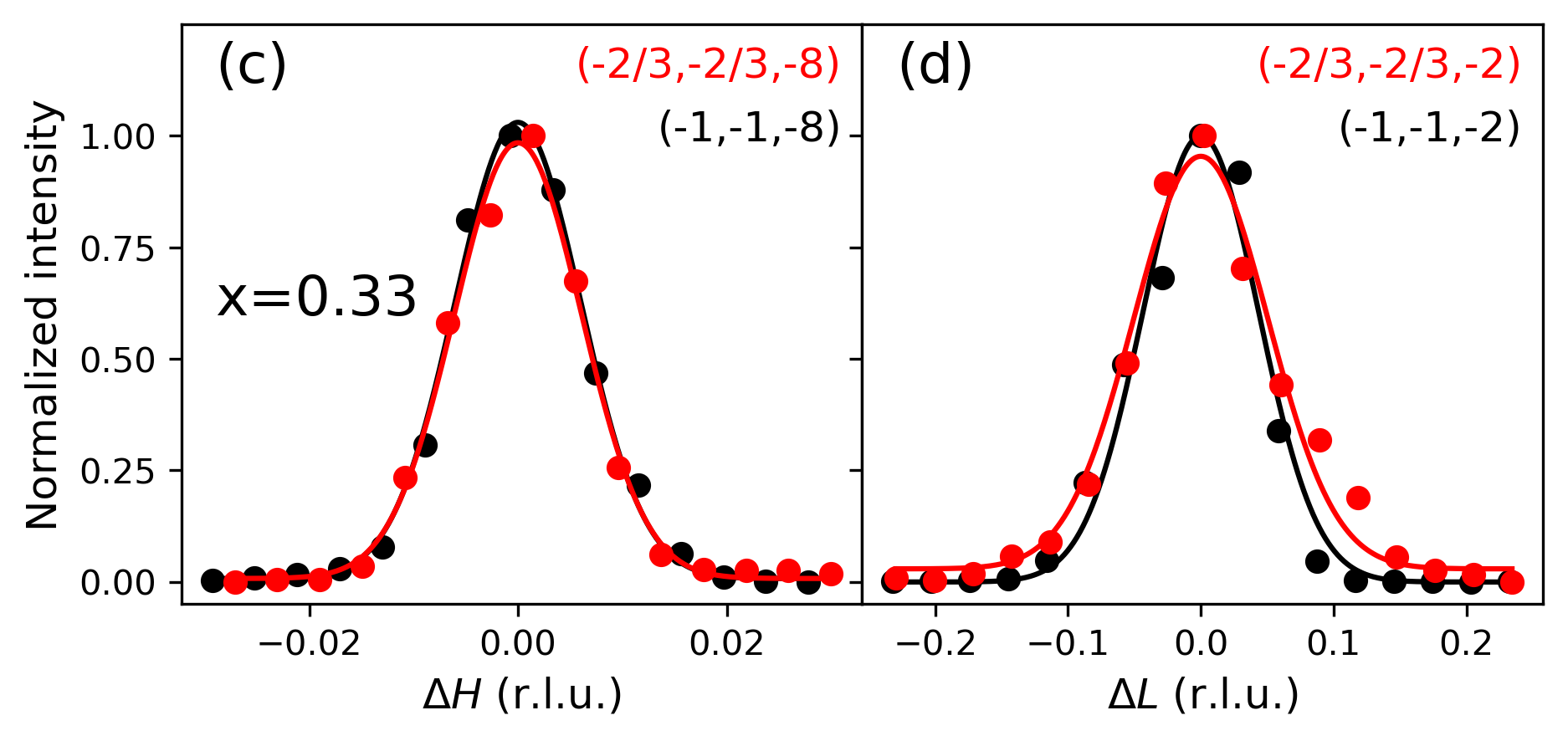}
	\caption{Representative single-crystal cuts through host (black) and superstructure (red) reflections of (a,b) Ni$_{1/4}$NbS$_2$ and (c,d) Ni$_{1/3}$NbS$_2$, along the in-plane direction (a,c) and along $c^*$ (b,d), centered on the fitted peak positions. The reflections, whose indices are given in each panel, were chosen among those whose widths are closest to the family averages of Table~\ref{Tab:Width}. In-plane cuts are along $(H,0,0)$ for $x=1/4$ and along $(H,H,0)$ for $x=1/3$. Lines are fits with a Gaussian plus a constant.}
\label{Fig:Widths}
\end{figure}

The widths of the host and superstructure reflections of the two commensurate compositions are compared in Table~\ref{Tab:Width} for the powders and for the single crystals, and representative single-crystal cuts are shown in Fig.~\ref{Fig:Widths}. Rather than converting them into correlation lengths, which would require a deconvolution of the very different resolution functions of the two setups, we compare the widths $\Delta q$ and use the ratio of superstructure to host widths within each technique as the relevant quantity.

In the single crystals, the superstructure reflections are as narrow as those of the host in both directions: the ratio $\Delta q^{\rm SS}/\Delta q^{\rm host}$ is 0.85(3) along $H$ and 0.87(7) along $L$ for $x=1/4$, and 0.98(6) and 1.08(6) for $x=1/3$. The Ni sublattice is therefore ordered over the same length scale as the NbS$_2$ host, within the resolution of the measurement, which is set by the pixel size of the reconstructions and by the mosaicity of the crystals.

\begin{table}[tbp]
\centering
\caption{Widths $\Delta q$ (in $10^{-3}$~\AA$^{-1}$) of the host and superstructure (SS) reflections along the in-plane ($H$) and out-of-plane ($L$) directions for Ni$_{0.25}$NbS$_2$ and Ni$_{0.33}$NbS$_2$ powders and single crystals. Powder: Scherrer widths, $\Delta q = 2\pi/\xi$ with $\xi = K\lambda/(\beta\cos\theta)$, $K = 0.9$ and $\beta$ the FWHM in radians, of the $2\bar{2}0$ and $002$ (host) and $0\frac12 0$ and $0\frac12 2$ (SS) reflections for $x=0.25$, and of the $2\bar{2}0$ and $002$ (host) and $\frac13\frac13 0$ and $\frac13\frac13 1$ (SS) reflections for $x=0.33$. Single crystal: average over all well-resolved reflections of each family (Sec.~\ref{sec:methods} C). Uncertainties are the fit errors for the powder and the standard errors of the mean for the crystals; the errors on the ratios are propagated from those of the widths. Absolute widths cannot be compared between the two setups.}
\label{Tab:Width}
\begin{ruledtabular}
\begin{tabular}{lcccc}
 & \multicolumn{2}{c}{Ni$_{0.25}$NbS$_2$} & \multicolumn{2}{c}{Ni$_{0.33}$NbS$_2$} \\
 & powder & crystal & powder & crystal \\
\hline
$\Delta q_H^{\text{host}}$ & 5.64(1)  & 48(1)   & 7.03(4)  & 56(3) \\
$\Delta q_H^{\text{SS}}$   & 17.70(5) & 41(1)   & 12.74(8) & 55(1) \\
SS/host                    & 3.14(1)  & 0.85(3) & 1.81(2)  & 0.98(6) \\
\hline
$\Delta q_L^{\text{host}}$ & 4.61(1)  & 78(3)   & 4.67(1)  & 56(3) \\
$\Delta q_L^{\text{SS}}$   & 53.3(5)  & 68(4)   & 16.49(4) & 61(1) \\
SS/host                    & 11.5(1)  & 0.87(7) & 3.53(1)  & 1.08(6) \\
\end{tabular}
\end{ruledtabular}
\end{table}

\begin{figure}[tbp]
	\centering
	\includegraphics[width=0.85\linewidth]{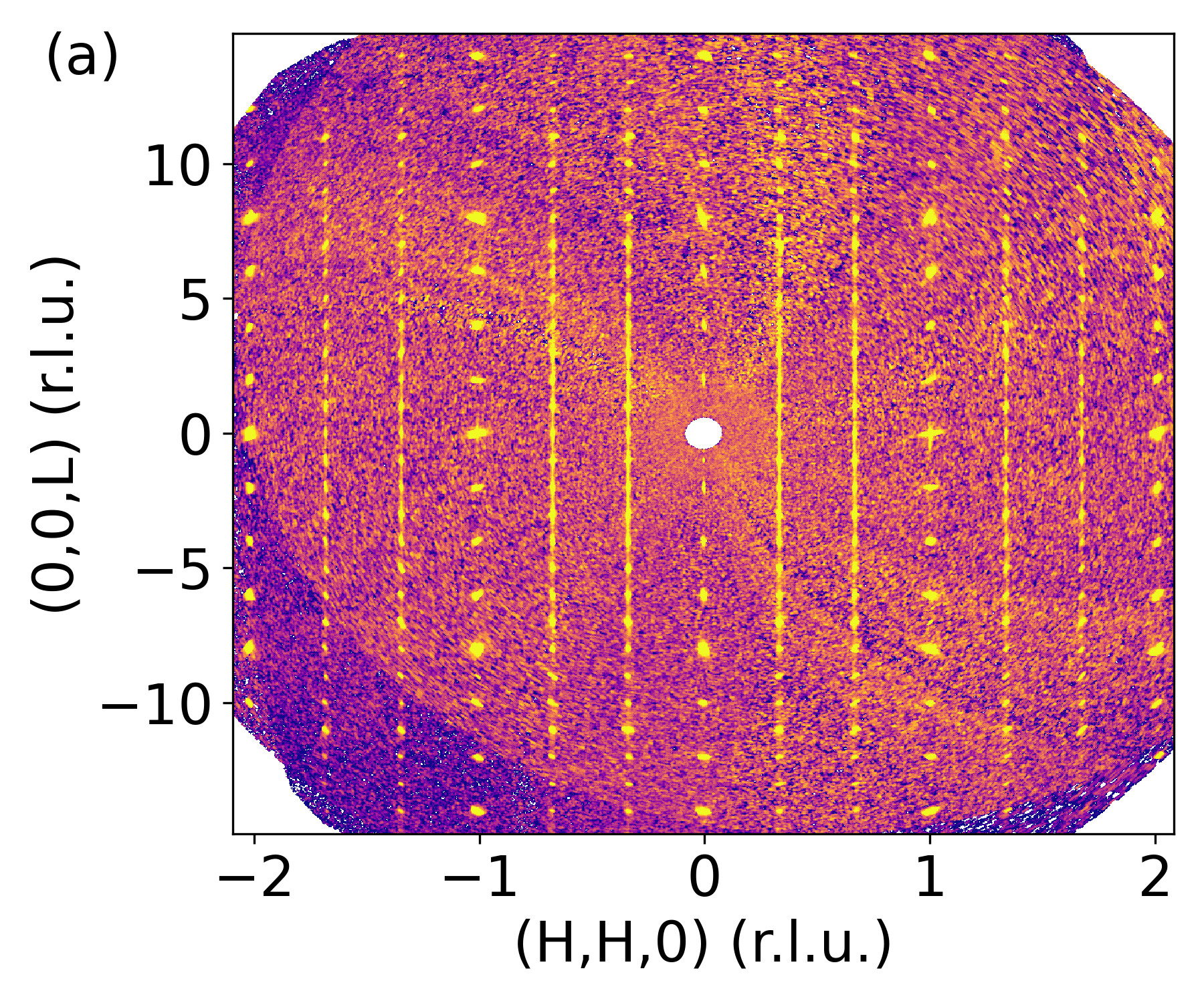}
	\includegraphics[width=0.85\linewidth]{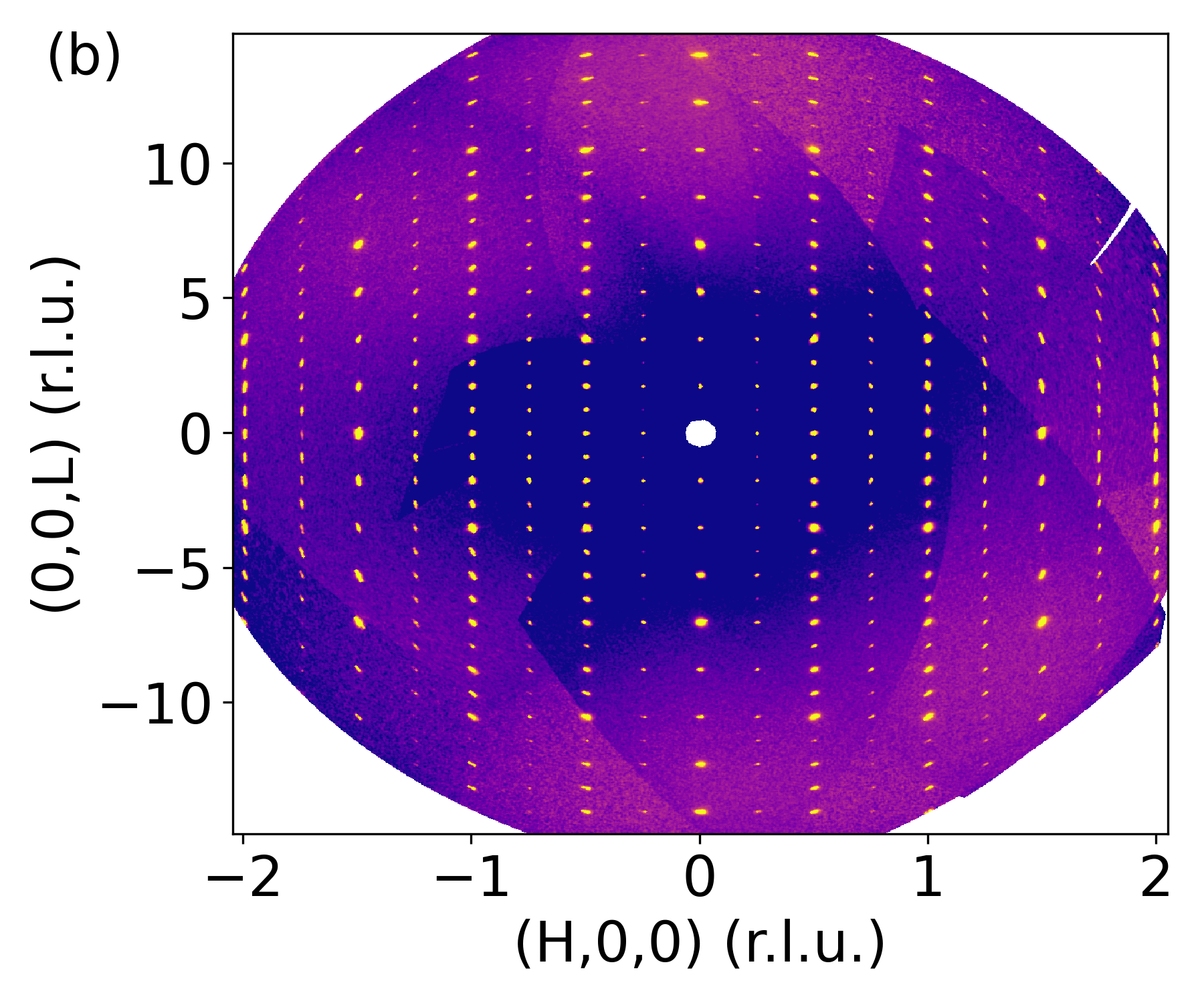}
	\caption{(a) Reciprocal-space reconstruction of the $(H,H,L)$ plane of Ni$_{1/3}$NbS$_2$, showing rods of diffuse scattering along $L$ at the superstructure positions $H = \pm\tfrac13, \pm\tfrac23, \dots$, while the host reflections at integer $H$ (allowed for even $L$ only) show no diffuse component. (b) Reconstruction of the $(H,0,L)$ plane of Ni$_{1/4}$NbS$_2$: neither the host reflections at integer $H$ nor the superstructure reflections at half-integer $H$ show a diffuse component along $L$.}
\label{Fig:Diffuse}
\end{figure}

Superimposed on the Bragg reflections, the reconstructions of Ni$_{1/3}$NbS$_2$ in the $(H,H,L)$ plane reveal continuous diffuse scattering (Fig.~\ref{Fig:Diffuse}). This diffuse intensity has two distinctive features: it forms rods along $c^*$ rather than isotropic clouds, and it is present only at the superstructure positions $H = \pm\tfrac13, \pm\tfrac23, \dots$, not at the host positions. Both point to a stacking disorder that affects the intercalant sublattice alone: the NbS$_2$ layers remain well stacked, while the registry of the Ni order from one gap to the next is only partially correlated. The coexistence of Bragg peaks and diffuse rods indicates that regions in which the $AB$ stacking of the Ni layers is maintained over many unit cells coexist with faulted regions. This is consistent with the short correlation length of the Ni order along $c$ found in the powders.

No such diffuse scattering is observed for $x=1/4$ [Fig.~\ref{Fig:Diffuse}(b)]. Between the Bragg reflections, the intensity along the superstructure rows $H=\pm\tfrac12, \pm\tfrac32$ of the $(H,0,L)$ plane exceeds the background measured halfway between rows by at most 20\%, as along the host rows, whereas along the superstructure rows of $x=1/3$ it reaches 1.7 to 2.3 times the background. The Ni layers of the $x=1/4$ crystal are therefore stacked without detectable faults, in line with the description of its structure by a single Ni site, whereas at $x=1/3$ the stacking faults accompany the partial occupation of the $2b$ site and the inversion twinning (Sec.~\ref{sec:chiral}).

\section{Discussion}
\label{sec:discussion}

\subsection{Single crystals versus powders}
\label{sec:SCvsP}

In contrast with the single crystals, the superstructure reflections of the powders, obtained by grinding crystals of the same batches, are markedly broader than those of the host: the ratio reaches 3.1 along $H$ and 11.5 along $L$ for $x=1/4$, and 1.8 and 3.5 for $x=1/3$ (Table~\ref{Tab:Width}), while the host reflections themselves remain narrow ($\Delta q \simeq 5$--$7\times10^{-3}$~\AA$^{-1}$). Grinding therefore degrades the Ni order much more strongly than the host lattice, and does so predominantly along $c$, i.e.\ by disrupting the stacking correlations of the intercalant layers rather than their in-plane order. This is consistent with the layered morphology of the crystals, which cleave along $ab$: the mechanical damage introduced by grinding acts primarily on the van der Waals gaps, which is precisely where the Ni sublattice resides. The effect is strongest at $x=1/4$, which is also the composition at which the lattice parameters of the two kinds of samples differ most.

Expressed in the parent hexagonal cell, the single crystals give $a = 3.31449(4)$~\AA\ and $c = 11.82484(16)$~\AA\ for $x=1/4$, and $a = 3.31757(6)$~\AA\ and $c = 11.8657(2)$~\AA\ for $x=1/3$. The lattice expands between the two compositions, as in the powders ($\Delta a = +0.0031$ and $\Delta c = +0.041$~\AA\ for the single crystals, against $+0.0029$ and $+0.028$~\AA\ for the powders). The single-crystal values of $a$ at both compositions and of $c$ at $x=1/3$ are smaller than the powder values by the same amount, 0.22--0.24\%, which we attribute to the wavelength calibration of the two setups. The value of $c$ at $x=1/4$ is instead 0.34\% smaller than in the powder, i.e.\ about 0.013~\AA\ below what the calibration alone would give. The single crystal, whose superstructure reflections are as narrow as the host ones (Sec.~\ref{sec:widths}) and which shows no diffuse scattering (Sec.~\ref{sec:diffuse}), therefore has a markedly shorter interlayer spacing than the powder, in which the Ni order is degraded by grinding: the minimum of $c$ at $x=1/4$ is even deeper in the best-ordered material.

\subsection{Distribution of Ni over the octahedral sites}
\label{sec:distribution}
\label{sec:chiral}
\label{sec:phasediag}

The single-crystal refinements and the superlattice intensities of the powders can both be read in terms of the distribution of Ni over the three octahedral sublattices of a van der Waals gap. In the $\sqrt3\times\sqrt3$ structure, Ni occupies one of them in each gap, the $2d$ site; a second one, $2c$, is the sublattice occupied in the neighboring gaps, and the third one, $2b$, is empty in all gaps. For the $\tfrac13\tfrac13\ell$ reflections the three sublattices carry the phases $1$, $\omega$ and $\omega^2$, with $\omega=e^{2\pi i/3}$, and the contributions of the two gaps of the cell add for even $\ell$ and subtract for odd $\ell$:
\begin{equation}
F_{\rm Ni}\propto S_1+(-1)^{\ell}S_2,\qquad S_j=n_j^{(1)}+\omega\,n_j^{(\omega)}+\omega^2 n_j^{(\omega^2)},
\label{eq:layers}
\end{equation}
where $n_j$ are the occupancies of the three sublattices in gap $j$. For the ideal $AB$ stacking, $S_1=1$ and $S_2=\omega$, and the even- and odd-$\ell$ reflections are in the ratio $|1+\omega|^2:|1-\omega|^2=1:3$. Ni on the $2b$ site, which belongs to the same sublattice in both gaps, adds the same term to $S_1$ and $S_2$: it cancels in the odd-$\ell$ reflections and reduces the even-$\ell$ ones, which vanish when the $2b$ occupancy reaches one half. Ni on the $2c$ sites has the opposite effect, while a random distribution over all vacant sites reduces both families alike, and stacking faults broaden both. Full structure-factor calculations based on the refined $x=1/3$ structure (Table~S3 of the SM~\cite{SM}) confirm this picture: placing the excess Ni of $x=0.46$ on the $2b$ site reduces the ratio $I(\tfrac13\tfrac13 0)/I(\tfrac13\tfrac13 1)$ by a factor of about 20 and slightly increases $I(\tfrac13\tfrac13 1)$ relative to the $002$ reflection, whereas a random distribution over $2b$ and $2c$ leaves the ratio unchanged.

This framework accounts for the intrinsic disorder found at $x=1/3$ and for its evolution at higher Ni content. The single-crystal analysis reveals an intercalant disorder that the Rietveld refinements cannot detect: at $x=1/3$, 9.0(2)\% of the Ni sits on the $2b$ site that the ideal $\sqrt3\times\sqrt3$ order leaves empty, whereas the $x=1/4$ data are described by a model with Ni on the $2a$ site only and show no diffuse scattering. Even the best-ordered $\sqrt3\times\sqrt3$ crystals therefore carry a few percent of quenched intercalant disorder, which acts directly on the magnetic lattice of these frustrated triangular systems~\cite{an2023,xie2022}. In this crystal, the transfer of 9\% of the Ni from $2d$ to $2b$ reduces the calculated even-$\ell$ intensities by about one half and the odd-$\ell$ ones by only 18\%.

Above $x=1/3$, the excess Ni must occupy sites that are vacant in the ideal structure. At $x=0.46$, the odd-$\ell$ reflection keeps its intensity while the even-$\ell$ one drops to 2\% (Sec.~\ref{sec:powder}), which is the signature of Ni on the $2b$ site. The ratio $I(\tfrac13\tfrac13 0)/I(\tfrac13\tfrac13 1)$ decreases by a factor of 33 with respect to $x=0.33$, close to the factor of 20 calculated when all the excess Ni is placed on $2b$, and corresponds to a $2b$ occupancy of about 0.4. A random distribution over the vacant sites would leave this ratio unchanged. Stacking faults cannot account for it either: random reversals of the $AB$ sequence broaden the even-$\ell$ reflections three times more than the odd-$\ell$ ones, whereas the widths of both reflections, and the correlation length along $c$ (Table~S1 of the SM~\cite{SM}), are the same at $x=0.46$ as at $x=0.33$. The excess Ni is thus accommodated on the $2b$ sublattice, precisely the minority site found in the $x=1/3$ crystal, with an average occupancy of about 0.4 if all of it is placed there. These Ni ions, about a quarter of all Ni at $x=0.46$, do not order among themselves over long distances, since they produce no superlattice reflection of their own. At $x=0.37$, the ratio is reduced almost as much, by a factor of 23, so that the $\sqrt3\times\sqrt3$ regions present in this sample already have a $2b$-rich configuration; but the odd-$\ell$ intensity, 60\% of that at $x=0.33$ where the $2b$-rich configuration would give 125\%, shows that these regions contain only about half of the Ni, the rest having no long-range order apart from weak $2\times2$ correlations. The range $0.37\le x\le0.46$ is therefore not a simple coexistence of two ordered phases, but a largely disordered state in which $\sqrt3\times\sqrt3$ correlations dominate and $2\times2$ correlations remain minor.

The stacking disorder of the Ni layers revealed by the diffuse rods is not independent of the chirality of the structure. In $P6_322$ it is precisely the $AB$ stacking of the Ni layers along $c$ that defines the handedness of the structure, so that a stacking fault is a natural boundary between domains of opposite chirality. The Flack parameter of 0.39(7) shows that such inverted domains are present in the very crystal that produces the diffuse rods, and the two observations are best regarded as two views of the same defect: a network of $c$-axis stacking faults separating chiral domains, of the kind imaged by transmission electron microscopy in Fe$_x$TaS$_2$~\cite{horibe2014} and consistent with the streaking of superlattice reflections along $c^*$ reported for Fe$_{1/4}$TaS$_2$~\cite{morosan2007}. Since the magnetism of Ni$_{1/3}$NbS$_2$ and of its Co and Cr analogues is discussed in terms of chiral and non-coplanar spin textures, the inversion-twin fraction, the stacking faults and the 9\% of Ni on the $2b$ site set a concrete structural bound on how clean a chiral magnetic ground state can be in such crystals.

The ordering of Ni in the van der Waals gap thus evolves non-monotonically with composition: disordered at $x=0.18$, $2\times2$ at $x=0.25$, $\sqrt3\times\sqrt3$ at $x=0.33$, a largely disordered state with dominant $\sqrt3\times\sqrt3$ correlations and weak $2\times2$ correlations at $x=0.37$--$0.46$, and a $2\times2$-type stripe order at $x=0.57$. This sequence is qualitatively consistent with cluster-expansion and Monte Carlo modeling of intercalant ordering in TaS$_2$~\cite{craig2025}, in which the $2\times2$ and $\sqrt3\times\sqrt3$ motifs compete and coexist away from their stoichiometric compositions as a result of intercalant--intercalant interactions acting on different length scales, and with the dual magnetic transitions reported for Fe$_{0.30}$NbSe$_2$~\cite{xu2026}. In the powders, the Ni order is well correlated in plane but much less along $c$, where its correlation length is about ten unit cells for $0.25 \le x \le 0.46$: the weakest link of the structure, and the one most sensitive to mechanical damage, is the registry of the Ni layers across the NbS$_2$ sheets. The NbS$_2$ host itself is pure 2H at all compositions except for a small 3R fraction at $x=0.18$, the lowest composition studied. This is compatible with the mixed 2H/3R stacking reported for $x \lesssim 0.15$~\cite{ramakrishna2025} and suggests that Ni intercalation stabilizes the 2H polytype.

\subsection{The $c$-axis minimum at $x=1/4$}
\label{sec:canomaly}

The minimum of the interlayer spacing at the commensurate composition $x=1/4$ cannot be accounted for by a simple steric (Vegard-type) effect, which would make $c$ grow monotonically with Ni content. Three observations bear on its origin.

First, the minimum is an intrinsic property of the ordered $2\times2$ structure. The relaxed DFT+$U$ supercells, which are perfectly ordered at every composition, reproduce a local minimum of $c$ at $x=1/4$, between $x=1/9$ and $x=1/3$ (Table~\ref{Tab:DFT}). The 9\% of misplaced Ni found at $x=1/3$ is therefore not required to explain why $c$ is larger at $x=1/3$ than at $x=1/4$.

Second, the comparison of the two refined structures (Fig.~S4 and Table~S2 of the SM~\cite{SM}) locates the contraction of $c/2$ between $x=1/3$ and $x=1/4$, $-20.4(1)$~m\AA. At $x=1/4$ the S layers bordering the gap split into two sites: S1 ($12k$, three S atoms in four), bonded to Ni, and S2 ($4f$, one in four), which has no Ni neighbor and caps the contracted Nb$_3$ triangle. S2 sits 0.050~\AA\ closer to the Ni plane than the S atoms at $x=1/3$, whereas S1 stays within 5~m\AA\ of its $x=1/3$ height, with a Ni--S bond of 2.3841(6)~\AA, slightly longer than the 2.3796(4)~\AA\ of the majority $2d$ site at $x=1/3$. The S2 displacement alone contributes $-25$~m\AA\ to $\Delta(c/2)$, through the van der Waals gap, while the small outward shift of S1 offsets $+7$~m\AA; the gap width, averaged over the S sites, thus decreases by $\Delta g = -18.0$~m\AA, whereas the S--Nb--S slab thickness changes by only $\Delta t = -2.4$~m\AA. The contraction of $c$ at $x=1/4$ is therefore not due to a shorter Ni--S bond but to the S atoms that are not bonded to Ni, which relax into the empty part of the gap above the Nb$_3$ trimers. DFT+$U$ reproduces the trimerization and the S2 shift ($-43$~m\AA), but moves S1 outward by 14~m\AA, which cancels the effect on $g$; this explains why the minimum of $c$ is weaker in the calculation than in experiment.

Third, the amplitude of the minimum is several times larger in experiment than in DFT (see above), and larger still in the best-ordered single crystal than in the powder (Sec.~\ref{sec:SCvsP}). The difference may come from the intercalant disorder that the calculation ignores: at $x=1/4$ the data are described by a single Ni site repeated in every gap, without detectable stacking faults, whereas at $x=1/3$ the partial occupation of the $2b$ site and the stacking faults of the Ni layers (Sec.~\ref{sec:diffuse}) make the bonding across the gaps less uniform; at $x=1/4$ itself, the degradation of the Ni stacking by grinding reduces the contraction of the powder with respect to the single crystal. A contraction of the stack is a collective response that requires every gap to be bonded in the same way, so that such disorder can only reduce it. The overall picture is thus that $c$ increases weakly with the Ni content, and that a minimum appears on top of this trend only where the Ni ions form a well-developed long-range order, as at $x=1/4$. X-ray absorption spectroscopy at the Ni $L_{2,3}$ and S $K$ edges at $x=1/4$ and $1/3$ would determine how the Ni $3d$--S $3p$ hybridization and the charge transfer to the host differ between the two structures, and in particular at the S2 sites that are not bonded to Ni.

\subsection{Stripe order at $x=0.57$}
\label{sec:disc057}
\label{sec:x057}

The powder pattern at $x=0.57$ differs from all others in two respects. At $x=0.57$ the host reflections with in-plane components are split. The $110$ and $220$ reflections, single and narrow at $x=0.25$, are each described by two components separated by $\Delta(2\theta) = 0.044^\circ$ and $0.100^\circ$, respectively (Fig.~S1 of the SM~\cite{SM}). Both splittings correspond to the same relative difference of interplanar spacings, $\Delta d/d = 1.9\times10^{-3}$, i.e.\ the splitting grows with the scattering angle, while the $00\ell$ reflections remain single. This is most naturally explained by an orthorhombic distortion of the hexagonal lattice. Indexing the hexagonal $hh0$ reflections in the orthohexagonal cell ($a_o = a$, $b_o \simeq \sqrt3\,a$), this corresponds to an orthorhombic strain $|b_o/\sqrt3 a_o - 1| \approx 2.5\times10^{-3}$. The $2\times2$-type order at $x=0.57$ is therefore not hexagonal, which points to a stripe arrangement of the Ni ions. Its superlattice reflections are narrower than at $x=0.25$, without measurable excess width along $c$ (SM~\cite{SM}): unlike the Ni layers at lower $x$, the stripes are correlated in three dimensions. This long correlation length should not be read as the signature of the best-ordered phase. A nearly half-filled stripe layer can be stacked on the next one in only a few distinct ways, so that configurational entropy hardly favors stacking faults, and the orthorhombic distortion couples the stripe orientation of successive layers; both favor a coherent stacking of the stripes whatever the amount of disorder within each layer, which other observations show to be large (see below).

The single-crystal structure of $x=1/4$ also provides the key to the reappearance of $2\times2$ reflections at $x=0.57$. In Ni$_{1/4}$NbS$_2$, Ni fills only the $2a$ column of the quadrupled cell, leaving three octahedral positions per van der Waals gap empty. Filling one of these columns, at $(\tfrac12,\tfrac12,0)$ and $(\tfrac12,\tfrac12,\tfrac12)$, brings the composition to $x=1/2$ without changing the $2a\times2a$ periodicity. The Ni ions then form chains of nearest neighbors along one of the three equivalent directions of the triangular lattice, separated by empty rows: the order becomes a stripe order, and selecting one direction lowers the symmetry from hexagonal to orthorhombic. The stripe order itself only requires the orthohexagonal $a\times\sqrt3a$ cell used in the DFT calculation (Table~\ref{Tab:DFT}). Keeping in addition the Nb trimerization of the $x=1/4$ structure, the corresponding structure has space group $Cmcm$ (No.~63) with $a_o = 2a = 6.6290$~\AA, $b_o = 2\sqrt3\,a = 11.4817$~\AA\ and $c_o = c$, i.e.\ the orthohexagonal setting of the same $2a\times2a$ lattice (Table~\ref{Tab:Cmcm}); a lower symmetry such as its subgroup $Pmma$ cannot be excluded from the present powder data. This model accounts naturally for the orthorhombic splitting of the host reflections: the relaxed DFT structure of the stripe order is indeed orthorhombic, with $b_o/\sqrt3 a_o - 1 = -1.0\%$, four times the strain measured at $x=0.57$, $|b_o/\sqrt3 a_o - 1| \approx 0.25\%$. The two values cannot be compared directly, since the calculation assumes a perfect stripe order in the minimal cell. The comparison is nevertheless instructive: at $x=1/4$ the measured structural response to the Ni order, the contraction of $c$, is larger than the calculated one, whereas at $x=0.57$ the measured orthorhombic strain is four times smaller. As the strain is measured within each orthorhombic domain, it is not reduced by the averaging over orientation variants but by the imperfection of the stripe order itself: it is expected to scale with the square of the stripe order parameter. The small strain thus points to a strongly imperfect stripe order. Moreover, since its lattice coincides with the $2a\times2a$ one, it produces superstructure reflections at the same half-integer positions of the parent hexagonal cell. The model corresponds to $x=1/2$, slightly less than the measured composition; the remaining $\simeq0.07$~Ni per formula unit, i.e.\ at least 12\% of the Ni, must be accommodated outside the stripe order. This is only a lower bound, and it already exceeds the disorder of the $x=1/3$ crystal, in which the misplaced Ni amounts to 9\% of the Ni but only 0.03 per formula unit.

\begin{table}[tbp]
  \centering
  \caption{Model $Cmcm$ (No.~63) structure of the $x=1/2$ stripe order, obtained from the refined Ni$_{1/4}$NbS$_2$ structure by additionally filling the $(\tfrac12,\tfrac12)$ octahedral column, with $a = 6.6290$~\AA, $b = 11.4817$~\AA, $c = 11.8248$~\AA\ and $Z = 16$, in the standard setting of the $C$-centered cell. The refined Ni occupancy 0.944 is kept on both Ni columns.}
  \label{Tab:Cmcm}
\begin{ruledtabular}
 \begin{tabular}{lcccc}
Atom & Wyckoff & $x$ & $y$ & $z$ \\ \hline
Ni1 & $4a$  & 0 & 0 & 0 \\
Ni2 & $4b$  & 1/2 & 0 & 0 \\
Nb1 & $4c$  & 0 & 0 & 1/4 \\
Nb2 & $4c$  & 1/2 & 0.99426 & 3/4 \\
Nb3 & $8g$  & 0.25861 & 0.24713 & 1/4 \\
S1  & $16h$ & 0.25146 & 0.91618 & 0.11897 \\
S2  & $8f$  & 1/2 & 0.16667 & 0.11436 \\
S3  & $8f$  & 1/2 & 0.66764 & 0.11897 \\
 \end{tabular}
 \end{ruledtabular}
\end{table}

The intensities calculated for this model are compared with those of the refined $x=1/4$ structure in Table~\ref{Tab:CalcInt}. Filling the $(\tfrac12,\tfrac12)$ column doubles the structure factor of the $\tfrac12\tfrac12\ell$ reflections of a single domain, multiplying their intensity by four, while it extinguishes the $\tfrac12 0\ell$ and $0\tfrac12\ell$ reflections, which the $x=1/4$ structure produces with comparable intensity. The three orientation variants of the stripes are equally probable, and averaging over them restores intensity at all half-integer positions. What survives the domain average is an overall enhancement of the superlattice
reflections relative to the 002 host reflection, by a factor 2.2 for $0\,\frac12\,0$
and 2.1 for $\frac12\,\frac12\,0$. Part of this enhancement comes from the reference
itself, since the 002 intensity drops by 38\,\% when the additional Ni is inserted,
the Ni and host contributions to this reflection being in antiphase; at a common
scale factor the two superlattice reflections gain 37\,\% and 31\,\%, respectively. The measured integrated intensities do not allow this prediction to be tested quantitatively (Table~S4 of the SM~\cite{SM}). Normalized to $002$, the $\tfrac12 00$ and $\tfrac12\tfrac12 0$ reflections are weaker at $x=0.57$ than at $x=0.25$ (by a factor of about 0.6), whereas normalized to the in-plane host reflection $100$, for which the model predicts almost no change (factors 0.95 and 0.91), they keep 75 and 80\% of their intensity at $x=0.25$. The $002$ reflection is the most sensitive to preferred orientation, and the $x=0.25$ powder is itself degraded by grinding (Sec.~\ref{sec:SCvsP}). The half-integer reflections of $x=0.57$ are thus not stronger than at $x=0.25$, but narrower, and their intensities are compatible with the stripe model without providing a measure of its degree of order. The $0\tfrac12 1$ reflection provides a useful internal reference: the two Ni layers of the cell are related by the $6_3$ screw axis, so that their contributions cancel exactly for odd $\ell$ in both models, and this reflection measures only the distortion of the Nb--S framework.

The same model does not account for the interlayer spacing: $x=0.57$ shows no contraction of $c$, whereas the perfectly ordered stripe structure at $x=1/2$ has, in DFT, an even shorter $c$ than at $x=1/4$ (Table~\ref{Tab:DFT}). First, the measured composition departs from $x=1/2$, so that at least 0.07~Ni per formula unit, 12\% of the Ni, must occupy defect positions. Second, the stripe order itself is far from perfect in our sample, as shown by its small orthorhombic strain. In both cases the bonding across the gaps is no longer uniform and the contraction is lost, as discussed for $x=1/4$ in Sec.~\ref{sec:canomaly}. The calculation of Table~\ref{Tab:CalcInt} keeps the Nb and S positions of the $x=1/4$ structure, whose Nb trimerization will reorganize once the second Ni column is filled; the predicted intensities are therefore robust for the reflections dominated by the Ni sublattice but only indicative for those dominated by the host. A Rietveld refinement of the superlattice intensities of the $x=0.57$ pattern and a single-crystal study at this composition, which would also allow the selective extinctions to be observed within single domains, are the natural next steps.

\begin{table}[tbp]
  \centering
  \caption{Calculated intensities of the $002$ host reflection and of three $2\times2$ superlattice reflections, indexed in the parent hexagonal cell, for the refined Ni$_{1/4}$NbS$_2$ structure ($P6_3/mmc$, Table~\ref{Tab:Refinement0p25}) and for the $x=1/2$ stripe model ($Cmcm$, Table~\ref{Tab:Cmcm}). Structure factors include Cromer--Mann atomic form factors and the refined displacement parameters; powder intensities $I \propto m\,\mathrm{LP}\,|F|^2$ are calculated for $\lambda = 0.6779$~\AA\ and normalized to $I(002) = 100$ for each model. For $x=1/2$, the value in parentheses is that of a single orthorhombic domain, the other one being the average over the three orientation variants.}
  \label{Tab:CalcInt}
    \begin{tabular}{lcc|cc}
      \hline
      \multirow{2}{*}{$hk\ell$} & \multirow{2}{*}{$d$ (\AA)} & \multirow{2}{*}{$m$} & $x=1/4$ & $x=1/2$ \\
       & & & $P6_3/mmc$ & $Cmcm$ \\
      \hline
      \multicolumn{5}{l}{\emph{Structure factors} $|F|^2$ (arb. units)}\\
      $002$                          & 5.912 & 2  & 55700 & 34800 \\
      $0\,\frac{1}{2}\,0$            & 5.741 & 6  & 2300  & 3160 \ (0) \\
      $0\,\frac{1}{2}\,1$            & 5.164 & 12 & 87    & 87 \ (87) \\
      $\frac{1}{2}\,\frac{1}{2}\,0$  & 3.314 & 6  & 2050  & 2680 \ (8040) \\
      \hline
      \multicolumn{5}{l}{\emph{Powder intensities}, $I(002)=100$}\\
      $002$                          & 5.912 & 2  & 100   & 100 \\
      $0\,\frac{1}{2}\,0$            & 5.741 & 6  & 11.7  & 25.7 \ (0) \\
      $0\,\frac{1}{2}\,1$            & 5.164 & 12 & 0.71  & 1.14 \ (1.14) \\
      $\frac{1}{2}\,\frac{1}{2}\,0$  & 3.314 & 6  & 3.43  & 7.19 \ (21.6) \\
      \hline
    \end{tabular}
\end{table}

\subsection{Implications for altermagnetism}
\label{sec:altermag}

Intercalated TMDs have recently emerged as a family of layered altermagnets. In the $2\times2$ ($P6_3/mmc$) and $\sqrt3\times\sqrt3$ ($P6_322$) structures, the two intercalant planes of the cell are related by the $6_3$ screw axis and not by a translation or an inversion, so that an A-type antiferromagnetic order, with ferromagnetic intercalant planes coupled antiferromagnetically along $c$, lifts the spin degeneracy of the bands without spin--orbit coupling~\cite{regmi2025altermagnetism,dayroberts2026}. Such a nonrelativistic spin splitting has been observed by photoemission in Co$_{1/4}$NbSe$_2$~\cite{regmi2025altermagnetism,dale2024,devita2025} and Co$_{1/4}$TaSe$_2$~\cite{sprague2026}, and first-principles surveys predict further candidates at $x=1/4$ and $x=1/3$ among the Fe, Co and Ni compounds~\cite{dayroberts2026,sah2026}. Because the effect rests on the symmetry operations that relate the magnetic planes of successive van der Waals gaps, it is directly sensitive to the structural features identified here.

At $x=1/4$, the Ni order is that assumed in these calculations: a single Ni site, $AA$-stacked from gap to gap, without detectable stacking faults. Two refinements are, however, needed. First, the Nb atoms not lying below Ni form a kagome network, invoked to explain the flat bands of the $x=1/4$ compounds~\cite{sah2026}; our refinement shows that this network is a breathing kagome, with a bond alternation of 3.2005 and 3.4285~\AA. Second, the ratio of the interlayer to the intralayer intercalant distance, $c/2a'$, has been proposed as an empirical descriptor of the magnetic stacking, values below 0.91 favoring the A-type antiferromagnetic order~\cite{sah2026}. In our $x=1/4$ crystal $c/2a'=0.892$, lowered by the contraction of $c$ at this composition, which would place Ni$_{1/4}$NbS$_2$ on the antiferromagnetic side, whereas first-principles calculations predict a ferromagnetic ground state for this compound~\cite{dayroberts2026}; magnetic measurements on the present crystals will decide between the two.

At $x=1/3$, the stacking faults of the Ni layers locally change the operation that relates neighboring Ni planes, the inversion twins juxtapose domains of opposite handedness, and 9\% of the Ni sits on the $2b$ site, which belongs to the same sublattice in both gaps. Above $x=1/3$, a large fraction of the Ni is disordered. Whether Ni$_x$NbS$_2$ hosts an altermagnetic phase thus depends not only on the magnetic ground state of the ideal ordered structures, but also on the degree of Ni order actually realized, which is best at $x=1/4$.

\section{Conclusion}

We have carried out a composition-resolved structural study of Ni$_x$NbS$_2$ for $0.18 \le x \le 0.57$, with Ni contents determined by EDX, combining synchrotron powder and single-crystal X-ray diffraction with DFT+$U$ calculations. All samples retain the 2H-NbS$_2$ host, with a small 3R fraction at $x=0.18$ only, while the ordering of Ni in the van der Waals gap follows a non-monotonic sequence: disorder at $x=0.18$, $2\times2$ order at $x=0.25$, $\sqrt3\times\sqrt3\,R30^\circ$ order at $x=0.33$, a largely disordered state at $x=0.37$--$0.46$, in which the excess Ni occupies the $2b$ sites left vacant by the $\sqrt3\times\sqrt3$ order, and an orthorhombic $2\times2$ stripe order at $x=0.57$. In all ordered phases except the stripe phase, the Ni order is much shorter-ranged along $c$ than in plane; in the stripe phase, the long-range stacking coexists with at least 12\% of the Ni outside the stripe sites and with an orthorhombic strain four times smaller than calculated, so that this phase is not the best-ordered one despite its long correlation length. The in-plane lattice parameter increases monotonically with $x$, while the interlayer spacing responds weakly, saturates above $x\approx0.37$, and shows a pronounced minimum at $x=0.25$.

Single-crystal refinements show that Ni$_{1/4}$NbS$_2$ is described by a centrosymmetric $2a\times2a$ structure ($P6_3/mmc$) with Ni on a single site and no diffuse scattering, accompanied by a $3.4\%$ trimerization of the Nb triangular net, whereas the chiral $\sqrt3a\times\sqrt3a$ order of Ni$_{1/3}$NbS$_2$ ($P6_322$) leaves the Nb sublattice undistorted but is intrinsically imperfect, with 9.0(2)\% of the Ni on the nominally vacant $2b$ site, a substantial inversion-twin fraction and stacking faults of the Ni layers. The contraction of $c$ at $x=1/4$ is carried by the S atoms that are not bonded to Ni and relax into the empty part of the van der Waals gap above the Nb$_3$ trimers, the Ni--S bond itself being slightly longer than at $x=1/3$. The relaxed DFT+$U$ structures reproduce a local minimum of $c$ at $x=1/4$, and an even shorter $c$ for the ordered stripe structure at $x=1/2$: a well-developed long-range Ni order contracts the interlayer spacing, whereas intercalant disorder suppresses this contraction.

Two continuations follow naturally. Magnetic and transport measurements on the same composition series will show whether the structural singularities identified here, the single-site order and contracted stacking at $x=1/4$, the intrinsic disorder at $x=1/3$ and the stripe order at $x=0.57$, have counterparts in the magnetic and electronic response, and in particular whether the A-type antiferromagnetic order required for altermagnetism is realized at $x=1/4$. On the structural side, single-crystal data at $x\approx0.57$, which would determine the stripe structure, and an analysis of the diffuse scattering of the disordered $x=0.18$ sample, which would reveal the short-range Ni correlations that precede the commensurate superstructures, would complete the picture.

The crystallographic data of Ni$_{1/4}$NbS$_2$ and Ni$_{1/3}$NbS$_2$ have been deposited at the Cambridge Crystallographic Data Centre under deposition numbers CCDC~2600597 and 2600598.

\section*{Acknowledgments}
We acknowledge SOLEIL for providing synchrotron beamtime on CRISTAL (proposal 20251118) and the Paris \^Ile-de-France Region in the framework of DIM MaTerRE (project DAC-VX). This work has been supported by the Croatian Science Foundation under the project numbers IP-2020-02-9666, IP-2025-02-7384. G. P. and P. P. acknowledge support of the project Cryogenic Centre at the Institute of Physics – KaCIF co-financed by the Croatian Government and the European Union through the European Regional Development Fund-Competitiveness and Cohesion Operational Programme (Grant No. KK.01.1.1.02.0012) and the project Ground states in competition – strong correlations, frustration and disorder — FrustKor financed by the Croatian Government and the European Union through the National Recovery and Resilience Plan 2021-2026 (NRRP).

\nocite{vdB68,Anz70,Luo17,Mandujano25,Wadge25,Wagner08,Zhu09,Park24,Wu22,Wu09,Cho26,morosan2007,Musfeldt2025}
\bibliography{biblio_v8}

@article{han2018,
  author = {Han, G. H. and Duong, D. L. and Keum, D. H. and Yun, S. J. and Lee, Y. H.},
  title = {van der {W}aals Metallic Transition Metal Dichalcogenides},
  journal = {Chem. Rev.},
  volume = {118},
  pages = {6297},
  year = {2018}
}

@article{naito1982,
  author = {Naito, M. and Tanaka, S.},
  title = {Electrical Transport Properties in {2H}-{NbS$_2$}, {-NbSe$_2$}, {-TaS$_2$} and {-TaSe$_2$}},
  journal = {J. Phys. Soc. Jpn.},
  volume = {51},
  pages = {219},
  year = {1982}
}

@article{vanmaaren1966,
  author = {van Maaren, M. H. and Schaeffer, G. M.},
  title = {Superconductivity in group {V}a dichalcogenides},
  journal = {Phys. Lett.},
  volume = {20},
  pages = {131},
  year = {1966}
}

@article{tissen2013,
  author = {Tissen, V. G. and Osorio, M. R. and Brison, J. P. and Nemes, N. M. and Garc\'ia-Hern\'andez, M. and Cario, L. and Rodi\`ere, P. and Vieira, S. and Suderow, H.},
  title = {Pressure dependence of superconducting critical temperature and upper critical field of {2H}-{NbS$_2$}},
  journal = {Phys. Rev. B},
  volume = {87},
  pages = {134502},
  year = {2013}
}

@article{regmi2025altermagnetism,
  author = {Regmi, R. B. and Bhandari, H. and Thapa, B. and Hao, Y. and Sharma, N. and McKenzie, J. and Chen, X. and Nayak, A. and El Gazzah, M. and M{\'a}rkus, B. G. and Forr{\'o}, L. and Liu, X. and Cao, H. and Mitchell, J. F. and Mazin, I. I. and Ghimire, N. J.},
  title = {Altermagnetism in the layered intercalated transition metal dichalcogenide {CoNb$_4$Se$_8$}},
  journal = {Nat. Commun.},
  volume = {16},
  pages = {4399},
  year = {2025}
}

@article{ubaldini2014,
  author = {Ubaldini, A. and Giannini, E.},
  title = {Improved chemical vapor transport growth of transition metal dichalcogenides},
  journal = {J. Cryst. Growth},
  volume = {401},
  pages = {878},
  year = {2014}
}

@article{doebelin2015,
  author = {Doebelin, N. and Kleeberg, R.},
  title = {Profex: a graphical user interface for the {R}ietveld refinement program {BGMN}},
  journal = {J. Appl. Crystallogr.},
  volume = {48},
  pages = {1573},
  year = {2015}
}

@article{friend1987,
  author = {Friend, R. H. and Yoffe, A. D.},
  title = {Electronic properties of intercalation complexes of the transition metal dichalcogenides},
  journal = {Adv. Phys.},
  volume = {36},
  pages = {1},
  year = {1987}
}

@article{nakayama2006,
  author = {Nakayama, M. and Miwa, K. and Ikuta, H. and Hinode, H. and Wakihara, M.},
  title = {Electronic Structure of Intercalation Compounds of {Co$_x$NbS$_2$}},
  journal = {Chem. Mater.},
  volume = {18},
  pages = {4996},
  year = {2006}
}

@article{craig2025,
  author = {Craig, I. M. and Kim, B. J. and Fender, S. S. and Letman, P. and Asker, S. M. and Limmer, D. T. and Bediako, D. K. and Griffin, S. M.},
  title = {Modeling the superlattice phase diagram of transition metal intercalation in bilayer {2H}-{TaS$_2$}},
  journal = {J. Am. Chem. Soc.},
  volume = {147},
  pages = {13629},
  year = {2025}
}

@article{an2023,
  title={Bulk properties of the chiral metallic triangular antiferromagnets Ni 1/3 Nb S 2 and Ni 1/3 Ta S 2},
  author={An, Yeochan and Park, Pyeongjae and Kim, Chaebin and Zhang, Kaixuan and Kim, Hyeoncheol and Avdeev, Maxim and Kim, Jaewon and Han, Myung-Joon and Noh, Han-Jin and Seong, Seungho and others},
  journal={Physical Review B},
  volume={108},
  number={5},
  pages={054418},
  year={2023},
  publisher={APS}
}

@article{popcevic2022,
  author = {Pop\v{c}evi\'c, P. and Utsumi, Y. and Bia{\l}o, I. and Tabis, W. and Gala, M. A. and Rosmus, M. and Kolodziej, J. J. and Tomaszewska, N. and Garb, M. and Berger, H. and Batisti\'c, I. and Bari\v{s}i\'c, N. and Forr\'o, L. and Tuti\v{s}, E.},
  title = {Role of intercalated cobalt in the electronic structure of {Co$_{1/3}$NbS$_2$}},
  journal = {Phys. Rev. B},
  volume = {105},
  pages = {155114},
  year = {2022}
}

@article{morosan2007,
  author = {Morosan, E. and Zandbergen, H. W. and Li, L. and Lee, M. and Checkelsky, J. G. and Heinrich, M. and Siegrist, T. and Ong, N. P. and Cava, R. J.},
  title = {Sharp switching of the magnetization in {Fe$_{1/4}$TaS$_2$}},
  journal = {Phys. Rev. B},
  volume = {75},
  pages = {104401},
  year = {2007}
}

@article{horibe2014,
  author = {Horibe, Y. and Yang, J. and Cho, Y.-H. and Luo, X. and Kim, S. B. and Oh, Y. S. and Huang, F.-T. and Asada, T. and Tanimura, M. and Jeong, D. and Cheong, S.-W.},
  title = {Color theorems, chiral domain topology, and magnetic properties of {Fe$_x$TaS$_2$}},
  journal = {J. Am. Chem. Soc.},
  volume = {136},
  pages = {8368},
  year = {2014}
}

@article{xie2022,
  author = {Xie, L. S. and Husremovi\'c, S. and Gonzalez, O. and Craig, I. M. and Bediako, D. K.},
  title = {Structure and magnetism of iron- and chromium-intercalated niobium and tantalum disulfides},
  journal = {J. Am. Chem. Soc.},
  volume = {144},
  pages = {9525},
  year = {2022}
}

@article{dolomanov2009,
  author = {Dolomanov, O. V. and Bourhis, L. J. and Gildea, R. J. and Howard, J. A. K. and Puschmann, H.},
  title = {{OLEX2}: a complete structure solution, refinement and analysis program},
  journal = {J. Appl. Crystallogr.},
  volume = {42},
  pages = {339},
  year = {2009}
}

@article{sheldrick2015,
  author = {Sheldrick, G. M.},
  title = {Crystal structure refinement with {SHELXL}},
  journal = {Acta Crystallogr. C},
  volume = {71},
  pages = {3},
  year = {2015}
}

@article{sheldrick2015a,
  author = {Sheldrick, G. M.},
  title = {{SHELXT} -- integrated space-group and crystal-structure determination},
  journal = {Acta Crystallogr. A},
  volume = {71},
  pages = {3},
  year = {2015}
}

@article{parsons2013,
  author = {Parsons, S. and Flack, H. D. and Wagner, T.},
  title = {Use of intensity quotients and differences in absolute structure refinement},
  journal = {Acta Crystallogr. B},
  volume = {69},
  pages = {249},
  year = {2013}
}

@article{wang_smartmat,
  author = {Wang, Z. and Li, R. and Su, C. and Loh, K. P.},
  title = {Intercalated phases of transition metal dichalcogenides},
  journal = {SmartMat},
  volume = {1},
  pages = {e1013},
  year = {2020}
}

@article{ramakrishna2025,
  author = {Ramakrishna, Nunavath and Sahoo, Subhasish and Agasti, Prasanjit and Burma, Aswin and Mishra, Aman and Rana, Rajanikanta and Samal, Saroj L.},
  title = {Intercalation Induced Structural Transition and Magnetic Properties in {Ni$_x$NbS$_2$}},
  journal = {Chem. Asian J.},
  volume = {20},
  number = {22},
  pages = {e00796},
  year = {2025},
  doi = {10.1002/asia.202500796}
}

@article{boucher2024,
  author = {Utsumi Boucher, Y. and Bia{\l}o, I. and Gala, M. A. and Tabi{\'s}, W. and Rosmus, M. and Olszowska, N. and Kolodziej, J. J. and Gudac, B. and Novak, M. and Chogondahalli Muniraju, N. K. and Batisti{\'c}, I. and Bari{\v s}i{\'c}, N. and Pop{\v c}evi{\'c}, P. and Tuti{\v s}, E.},
  title = {Intercalation-induced states at the {F}ermi level and the coupling of intercalated magnetic ions to conducting layers in {Ni$_{1/3}$NbS$_2$}},
  journal = {Phys. Rev. B},
  volume = {109},
  pages = {085135},
  year = {2024}
}

@article{xu2026,
  author = {Xu, X. and Jiang, B. and Wang, R. and Qiu, Z. and Guo, S. and Lv, B. and Zhong, R.},
  title = {Stoichiometry-controlled structural order and tunable antiferromagnetism in {Fe$_x$NbSe$_2$} ($0.05 \le x \le 0.38$)},
  journal = {Phys. Rev. B},
  volume = {113},
  pages = {134420},
  year = {2026}
}

@article{vdB68,
  author = {van den Berg, J. M. and Cossee, P.},
  title = {Structural aspects and magnetic behavior of {NbS$_2$} and {TaS$_2$} containing extra metal atoms of the first transition series},
  journal = {Inorg. Chim. Acta},
  volume = {2},
  pages = {143},
  year = {1968}
}

@article{Anz70,
  author = {Anzenhofer, K. and van den Berg, J. M. and Cossee, P. and Helle, J. N.},
  title = {The crystal structure and magnetic susceptibilities of {MnNb$_3$S$_6$}, {FeNb$_3$S$_6$}, {CoNb$_3$S$_6$} and {NiNb$_3$S$_6$}},
  journal = {J. Phys. Chem. Solids},
  volume = {31},
  pages = {1057},
  year = {1970}
}

@article{Luo17,
  author = {Luo, H. and Strychalska-Nowak, J. and Li, J. and Tao, J. and Klimczuk, T. and Cava, R. J.},
  title = {S-shaped suppression of the superconducting transition temperature in {Cu}-intercalated {NbSe$_2$}},
  journal = {Chem. Mater.},
  volume = {29},
  pages = {3704},
  year = {2017}
}

@article{Mandujano25,
  author = {Mandujano, H. C. and Zavalij, P. Y. and Manj{\'o}n-Sanz, A. and Cao, H. and Rodriguez, E. E.},
  title = {Coexistence of commensurate and incommensurate antiferromagnetic ground states in {Co$_x$NbSe$_2$} single crystal},
  journal = {Phys. Rev. Mater.},
  volume = {9},
  pages = {014403},
  year = {2025}
}

@article{Wadge25,
  title = {Magnetic frustration enforced electronic reconstruction in Ni-intercalated ${\mathrm{NbSe}}_{2}$: Suppression of electronic orders},
  author = {Wadge, Ashutosh S. and Kazakov, Alexander and Gong, Xujia and Jastrzebski, Daniel and Kowalski, Bogdan J. and Lynnyk, Artem and Plucinski, Lukasz and Fakhredine, Amar and Diduszko, Ryszard and Aleszkiewicz, Marta and Korczak, Jedrzej and Wutke, Dawid and Rosmus, Marcin and Kurleto, Rafal and Olszowska, Natalia and Autieri, Carmine and Wisniewski, Andrzej},
  journal = {Phys. Rev. B},
  volume = {112},
  issue = {23},
  pages = {235129},
  numpages = {10},
  year = {2025},
  month = {Dec},
  publisher = {American Physical Society},
  doi = {10.1103/g8s4-dnxf},
  url = {https://link.aps.org/doi/10.1103/g8s4-dnxf}
}

@article{Wagner08,
  author = {Wagner, K. E. and Morosan, E. and Hor, Y. S. and Tao, J. and Zhu, Y. and Sanders, T. and McQueen, T. M. and Zandbergen, H. W. and Williams, A. J. and West, D. V. and Cava, R. J.},
  title = {Tuning the charge density wave and superconductivity in {Cu$_x$TaS$_2$}},
  journal = {Phys. Rev. B},
  volume = {78},
  pages = {104520},
  year = {2008}
}

@article{Zhu09,
  author = {Zhu, X. D. and Sun, Y. P. and Zhang, S. B. and Lei, H. C. and Li, L. J. and Zhu, X. B. and Yang, Z. R. and Song, W. H. and Dai, J. M.},
  title = {Superconductivity and single crystal growth of {Ni$_{0.05}$TaS$_2$}},
  journal = {Solid State Commun.},
  volume = {149},
  pages = {1296},
  year = {2009},
  doi = {10.1016/j.ssc.2009.05.007}
}

@article{Park24,
  author = {Park, P. and Cho, W. and Kim, C. and An, Y. and Avdeev, M. and Iida, K. and Kajimoto, R. and Park, J.-G.},
  title = {Composition dependence of bulk properties in the {Co}-intercalated transition metal dichalcogenide {Co$_{1/3}$TaS$_2$}},
  journal = {Phys. Rev. B},
  volume = {109},
  pages = {L060403},
  year = {2024}
}

@article{Wu22,
  author = {Wu, S. and Xu, Z. and Haley, S. C. and Weber, S. F. and Acharya, A. and Maniv, E. and Qiu, Y. and Aczel, A. A. and Settineri, N. S. and Neaton, J. B. and Analytis, J. G. and Birgeneau, R. J.},
  title = {Highly tunable magnetic phases in transition-metal dichalcogenide {Fe$_{1/3+\delta}$NbS$_2$}},
  journal = {Phys. Rev. X},
  volume = {12},
  pages = {021003},
  year = {2022}
}

@article{Wu09,
  author = {Wu, G. and Wu, T. and Li, Z. and Zhao, L. and Liu, R. H. and Chen, H. and Fang, D. F. and Luo, J. L. and Chen, X. H.},
  title = {Transport properties and the large anisotropic magnetoresistance of {Cu$_x$NbS$_2$} single crystals},
  journal = {J. Phys.: Condens. Matter},
  volume = {21},
  pages = {275601},
  year = {2009}
}

@article{Cho26,
  author = {Cho, W. and Nam, K. and An, Y. and Kim, Y. Y. and Jung, M.-H. and Kim, K. H. and Park, J.-G.},
  title = {Composition-dependent bulk properties of intercalated transition metal dichalcogenides {Co$_{1/3(1\pm\delta)}$NbS$_2$}},
  journal = {arXiv:2603.26069},
  year = {2026}
}

@article{giannozzi2009,
  author = {Giannozzi, P. and others},
  title = {{QUANTUM ESPRESSO}: a modular and open-source software project for quantum simulations of materials},
  journal = {J. Phys.: Condens. Matter},
  volume = {21},
  pages = {395502},
  year = {2009}
}

@article{giannozzi2017,
  author = {Giannozzi, P. and others},
  title = {Advanced capabilities for materials modelling with {QUANTUM ESPRESSO}},
  journal = {J. Phys.: Condens. Matter},
  volume = {29},
  pages = {465901},
  year = {2017}
}

@article{perdew2008,
  author = {Perdew, J. P. and Ruzsinszky, A. and Csonka, G. I. and Vydrov, O. A. and Scuseria, G. E. and Constantin, L. A. and Zhou, X. and Burke, K.},
  title = {Restoring the density-gradient expansion for exchange in solids and surfaces},
  journal = {Phys. Rev. Lett.},
  volume = {100},
  pages = {136406},
  year = {2008}
}

@article{dalcorso2014,
  author = {Dal Corso, A.},
  title = {Pseudopotentials periodic table: From {H} to {Pu}},
  journal = {Comput. Mater. Sci.},
  volume = {95},
  pages = {337},
  year = {2014}
}

@article{marzari1999,
  author = {Marzari, N. and Vanderbilt, D. and De Vita, A. and Payne, M. C.},
  title = {Thermal contraction and disordering of the {Al}(110) surface},
  journal = {Phys. Rev. Lett.},
  volume = {82},
  pages = {3296},
  year = {1999}
}

@article{dudarev1998,
  author = {Dudarev, S. L. and Botton, G. A. and Savrasov, S. Y. and Humphreys, C. J. and Sutton, A. P.},
  title = {Electron-energy-loss spectra and the structural stability of nickel oxide: An {LSDA+U} study},
  journal = {Phys. Rev. B},
  volume = {57},
  pages = {1505},
  year = {1998}
}

@article{timrov2018,
  author = {Timrov, I. and Marzari, N. and Cococcioni, M.},
  title = {Hubbard parameters from density-functional perturbation theory},
  journal = {Phys. Rev. B},
  volume = {98},
  pages = {085127},
  year = {2018}
}

@article{timrov2021,
  author = {Timrov, I. and Marzari, N. and Cococcioni, M.},
  title = {Self-consistent {Hubbard} parameters from density-functional perturbation theory in the ultrasoft and projector-augmented wave formulations},
  journal = {Phys. Rev. B},
  volume = {103},
  pages = {045141},
  year = {2021}
}

@article{timrov2022,
  author = {Timrov, I. and Marzari, N. and Cococcioni, M.},
  title = {{HP} -- {A} code for the calculation of {Hubbard} parameters using density-functional perturbation theory},
  journal = {Comput. Phys. Commun.},
  volume = {279},
  pages = {108455},
  year = {2022}
}

@misc{SM,
  note = {See Supplemental Material at [URL will be inserted by publisher] for the orthorhombic splitting at $x=0.57$, the correlation lengths of the powders, a comparison of $c(x)$ with other intercalated transition metal dichalcogenides, the structural response to the $2\times2$ order at $x=1/4$, and examples of Rietveld refinements, and calculated superlattice intensities for different Ni distributions above $x=1/3$, which includes Refs.~\cite{vdB68,Anz70,Luo17,Mandujano25,Wadge25,Wagner08,Zhu09,Park24,Wu22,Wu09,Cho26,morosan2007,Musfeldt2025}}
}

@article{Musfeldt2025,
  author = {Musfeldt, J. L. and Yang, J. and Putnam, R. E. and Zhang, S. and Haraldsen, J. T. and Cheong, S. W. and Liu, Z.},
  title = {Intercalated chalcogenides {Fe$_{1/4}$TaS$_2$} and {Fe$_{1/3}$TaS$_2$} under extreme pressure--temperature conditions},
  journal = {npj 2D Mater. Appl.},
  volume = {10},
  pages = {3},
  year = {2026},
  doi = {10.1038/s41699-025-00637-2}
}

@article{smejkal2022a,
  author = {\v{S}mejkal, L. and Sinova, J. and Jungwirth, T.},
  title = {Beyond conventional ferromagnetism and antiferromagnetism: A phase with nonrelativistic spin and crystal rotation symmetry},
  journal = {Phys. Rev. X},
  volume = {12},
  pages = {031042},
  year = {2022}
}

@article{smejkal2022b,
  author = {\v{S}mejkal, L. and Sinova, J. and Jungwirth, T.},
  title = {Emerging research landscape of altermagnetism},
  journal = {Phys. Rev. X},
  volume = {12},
  pages = {040501},
  year = {2022}
}

@article{dale2024,
  author = {Dale, N. and Ashour, O. A. and Vila, M. and Regmi, R. B. and Fox, J. and Johnson, C. W. and Fedorov, A. and Stibor, A. and Ghimire, N. J. and Griffin, S. M.},
  title = {Relativistic and nonrelativistic spin splitting above and below the {F}ermi level in a $g$-wave altermagnet},
  journal = {arXiv:2411.18761},
  year = {2024}
}

@article{devita2025,
  author = {De Vita, A. and Bigi, C. and Romanin, D. and Watson, M. D. and Polewczyk, V. and Zonno, M. and Bertran, F. and others},
  title = {Robust spin splitting and fermiology in a layered altermagnet},
  journal = {arXiv:2502.20010},
  year = {2025}
}

@article{sprague2026,
  author = {Sprague, M. and Mondal, M. I. and Sakhya, A. P. and Regmi, R. B. and Sadhukhan, S. and Kumay, A. K. and Sheokand, H. and Mazin, I. I. and Ghimire, N. J. and Neupane, M.},
  title = {Observation of altermagnetic spin-splitting in an intercalated transition metal dichalcogenide},
  journal = {Nat. Commun.},
  volume = {17},
  pages = {9935},
  year = {2026},
  doi = {10.1038/s41467-026-76784-x}
}

@article{sah2026,
  author = {Sah, A. and Lim, T. Y. and Conner, C. and Chakraborty, A. and Vignale, G. and Chang, T.-R. and Sukhachov, P. and others},
  title = {Altermagnetism, kagome flat band, and {W}eyl fermion states in magnetically intercalated transition metal dichalcogenides},
  journal = {npj Quantum Mater.},
  volume = {11},
  pages = {63},
  year = {2026},
  doi = {10.1038/s41535-026-00885-5}
}

@article{dayroberts2026,
  author = {Day-Roberts, E. and Wu, H. and Erten, O. and Botana, A. S.},
  title = {An altermagnetic materials library in intercalated transition-metal dichalcogenides},
  journal = {arXiv:2601.02481},
  year = {2026}
}
\bibliographystyle{apsrev4-1}
\end{document}


\title{Supplemental Material for ``Ni ordering in the van der Waals gap of Ni-intercalated 2H-NbS$_2$: from disorder through $2\times2$ and $\sqrt{3}\times\sqrt{3}$ order to stripe order''}

\author{Naveen Singh Dhami}
\email{naveen.dhami@universite-paris-saclay.fr}
\affiliation{Universit\'e Paris-Saclay, CNRS, Laboratoire de Physique des Solides, 91405 Orsay, France}
\affiliation{Synchrotron SOLEIL, L'Orme des Merisiers, Saint Aubin BP 48, 91192 Gif-sur-Yvette, France}

\author{Ricardo Guillermo Torres Ram\'irez}
\affiliation{Synchrotron SOLEIL, L'Orme des Merisiers, Saint Aubin BP 48, 91192 Gif-sur-Yvette, France}

\author{Gaurav Pransu}
\affiliation{Institute of Physics, Bijeni\v{c}ka cesta 46, 10000 Zagreb, Croatia}

\author{Ivo Batisti\'c}
\affiliation{Department of Physics, Faculty of Science, University of Zagreb, Bijeni\v{c}ka 32, 10000 Zagreb, Croatia}

\author{Pierre Fertey}
\affiliation{Synchrotron SOLEIL, L'Orme des Merisiers, Saint Aubin BP 48, 91192 Gif-sur-Yvette, France}

\author{Erik Elkaim}
\affiliation{Synchrotron SOLEIL, L'Orme des Merisiers, Saint Aubin BP 48, 91192 Gif-sur-Yvette, France}

\author{El-Eulmi Bendeif}
\affiliation{Synchrotron SOLEIL, L'Orme des Merisiers, Saint Aubin BP 48, 91192 Gif-sur-Yvette, France}

\author{Bruno Gudac}
\affiliation{Institute of Physics, Bijeni\v{c}ka cesta 46, 10000 Zagreb, Croatia}

\author{Nikolina Penić}
\affiliation{Department of Physics, Faculty of Science, University of Zagreb, Bijeni\v{c}ka 32, 10000 Zagreb, Croatia}

\author{Naveen Kumar Chogondahalli Muniraju
}
\affiliation{The Henryk Niewodnicza\'nski Institute of Nuclear Physics, Polish Academy of Sciences, ul. Radzikowskiego 152, 31-342 Krak\'ow, Poland}

\author{Neven Bari\v{s}i\'c}
\affiliation{Department of Physics, Faculty of Science, University of Zagreb, Bijeni\v{c}ka 32, 10000 Zagreb, Croatia}
\affiliation{Institute of Solid State Physics, TU Wien, 1040 Vienna, Austria}

\author{Petar Pop\v{c}evi\'c}
\email{ppopcevic@ifs.hr}
\affiliation{Institute of Physics, Bijeni\v{c}ka cesta 46, 10000 Zagreb, Croatia}

\author{Victor Bal\'edent}
\email{victor.baledent@universite-paris-saclay.fr}
\affiliation{Universit\'e Paris-Saclay, CNRS, Laboratoire de Physique des Solides, 91405 Orsay, France}
\affiliation{Institut Universitaire de France (IUF), Paris, France}

\maketitle

\section{Orthorhombic splitting at $x=0.57$}

\begin{figure}[!htb]
    \centering
    \includegraphics[width=1.0\textwidth]{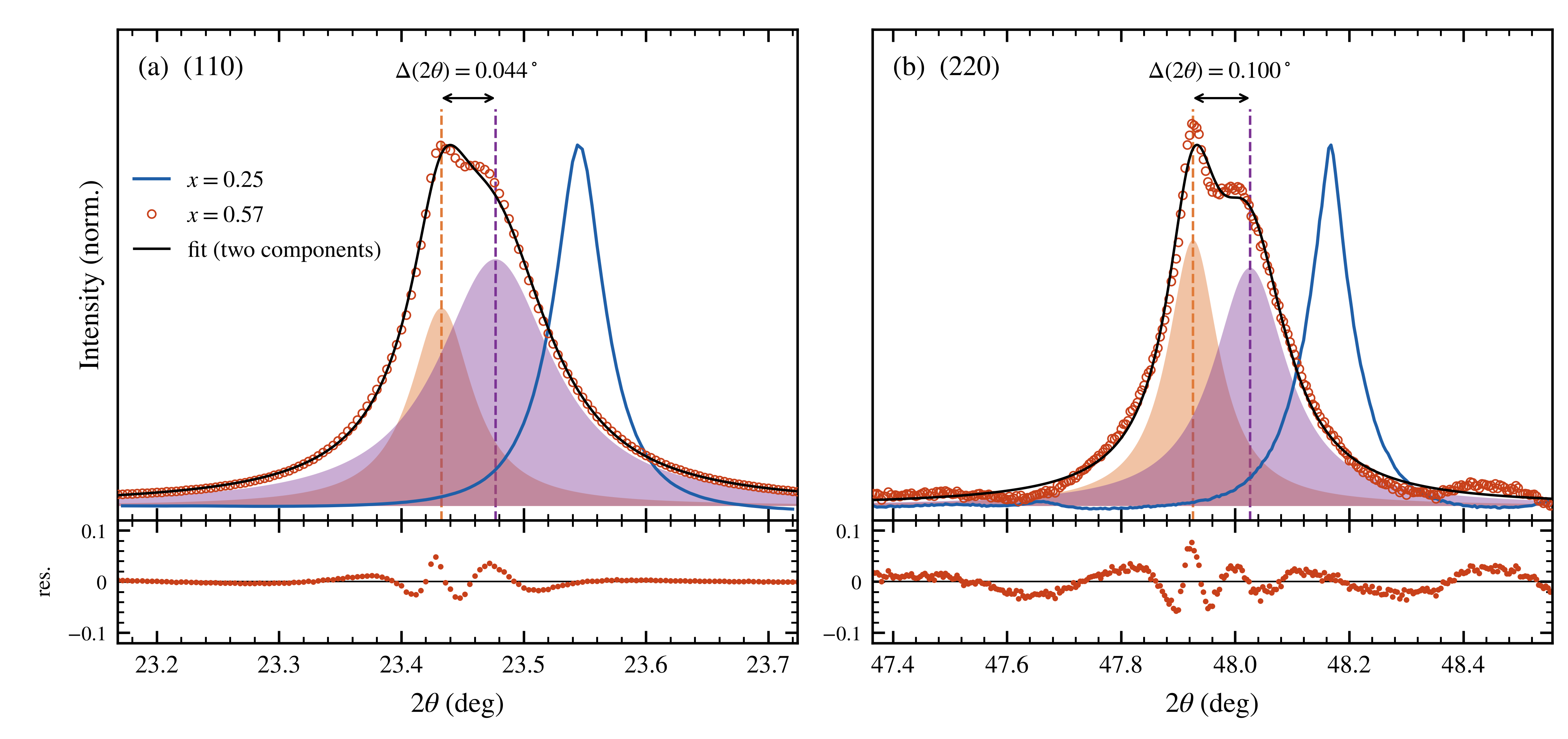}
    \caption{Splitting of the host reflections of Ni$_{0.57}$NbS$_2$. (a) $110$ and (b) $220$ reflections, indexed in the parent hexagonal cell, of Ni$_{0.57}$NbS$_2$ (open circles) compared with the single, narrow reflections of Ni$_{0.25}$NbS$_2$ (blue line). Both patterns are background-subtracted and normalized to their maximum. The black line is a fit with two pseudo-Voigt components (shaded), whose centers (dashed lines) are separated by $0.044^\circ$ in (a) and $0.100^\circ$ in (b), corresponding in both cases to $\Delta d/d = 1.9\times10^{-3}$. The splitting grows with the scattering angle, as expected for an orthorhombic distortion of the hexagonal lattice. The lower strips show the fit residuals.}
    \label{figOrtho}
\end{figure}

\section{Correlation lengths in the powders}

\begin{figure}[!htb]
    \centering
    \includegraphics[width=0.48\textwidth]{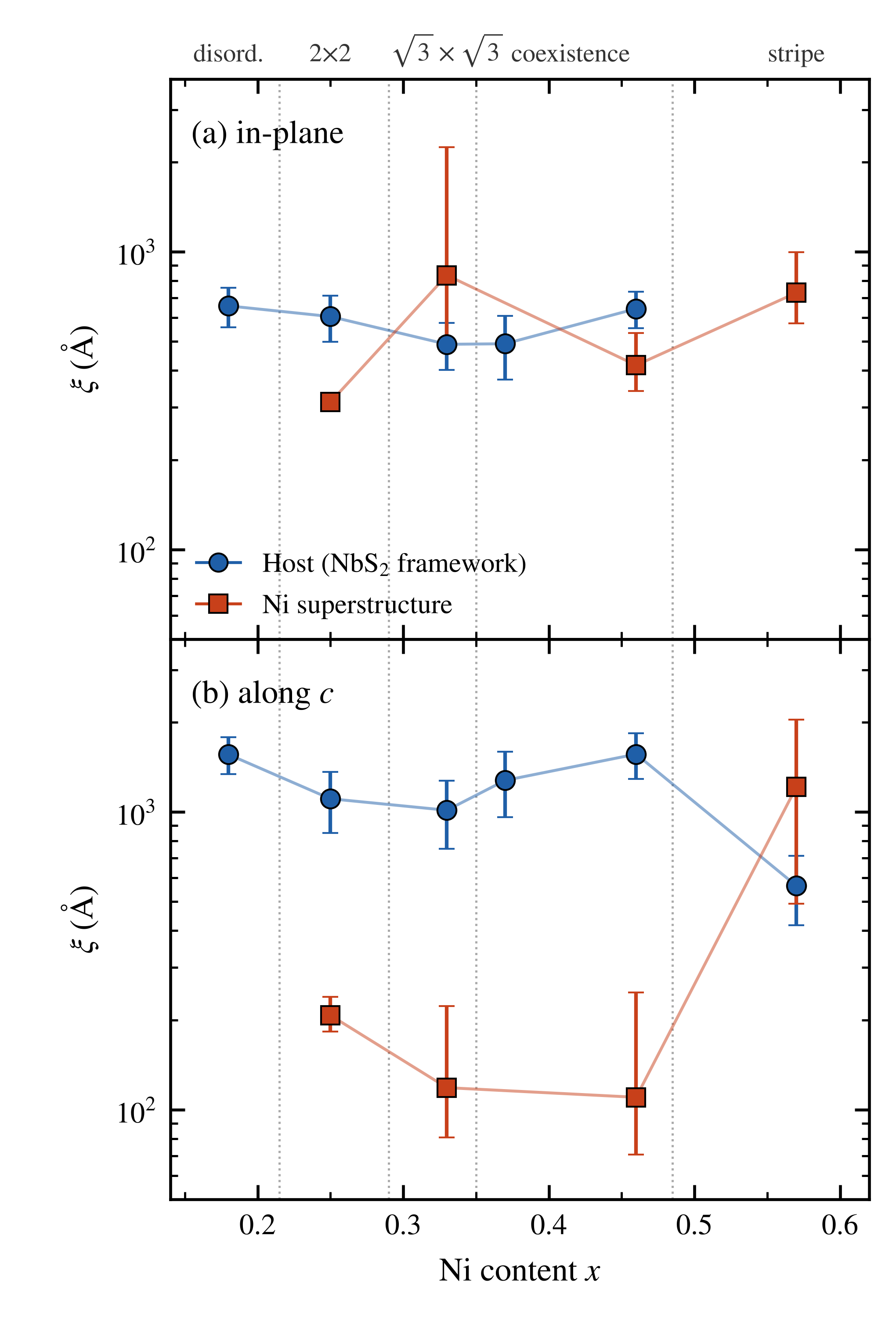}
    \caption{Correlation lengths of the NbS$_2$ host (circles) and of the Ni superstructure (squares) in Ni$_x$NbS$_2$ powders, (a) in plane and (b) along $c$, versus Ni content $x$ (EDX). Host: mean $\pm$ standard error over the accepted reflections. Superstructure: anisotropic fit to all resolved superlattice reflections after removal of the host width, with error bars from leave-one-reflection-out fits. Dotted lines indicate approximate boundaries between ordering regimes.}
    \label{figCorr}
\end{figure}

Figure~\ref{figCorr} and Table~\ref{tab:xi} give the correlation lengths of the NbS$_2$ host and of the Ni superstructure in the powders, in plane and along $c$, obtained from all resolved reflections. The host is well correlated in every sample: $\xi_{ab}\approx 490$--$660$~\AA\ with no significant dependence on $x$, and $\xi_{c}\approx 1000$--$1560$~\AA\ for $x=0.18$--$0.46$. The only host anomaly is a reduction of $\xi_{c}$ to about $565$~\AA\ at $x=0.57$, where the lattice becomes orthorhombic. The superstructure correlation lengths are defined differently: they are obtained after removing the host width at the same $q$, so that they measure the \emph{additional} disorder of the Ni sublattice with respect to the NbS$_2$ framework, and should not be compared directly with the host values.

In plane, the Ni order is correlated over several hundred \AA\ at all compositions where a superstructure is observed ($\xi_{ab}\approx 310$~\AA\ at $x=0.25$, $420$~\AA\ at $x=0.46$, and $730$~\AA\ at $x=0.57$; the value at $x=0.33$ is poorly constrained). Along $c$ the picture is very different. For $x=0.25$--$0.46$ the superstructure is correlated over only $\approx 100$--$200$~\AA, i.e.\ about ten unit cells, five to fourteen times less than the host in the same direction. The stacking of the Ni layers from one van der Waals gap to the next is therefore the weakest correlation in these samples, consistent with the diffuse rods along $c^{*}$ observed in the single-crystal data at $x=1/3$ (Sec.~IV\,B of the main text). The exception is $x=0.57$, where the superstructure shows no measurable excess width along $c$ ($\xi_{c}>500$~\AA), so that the Ni stripes are correlated in three dimensions. This long correlation length along $c$ does not imply that the stripe order is the best developed one: at least 12\% of the Ni lies outside the stripe sites and the orthorhombic strain is four times smaller than calculated (Sec.~V\,D of the main text).

At $x=0.57$ the $hk0$ host reflections split into orthorhombic doublets (Fig.~\ref{figOrtho}), so that no in-plane host width could be extracted, and the host reference used to subtract the instrumental and strain broadening at this composition is built from the $00\ell$ reflections alone; the superstructure values at $x=0.57$ should therefore be regarded as provisional. The widths quoted in Table~V of the main text are Scherrer widths of single reflections, without subtraction of the host width; they lead to the same conclusion, namely superstructure reflections much broader than the host reflections along $c$.

\begin{table}[!htb]
\caption{Correlation lengths $\xi=2\pi/\Delta q$ (\AA) shown in Fig.~\ref{figCorr}. Host: mean over $N$ instrument-corrected $hk0$ (in-plane) or $00\ell$ (along $c$) reflections, error = standard error of the mean ($N$ in parentheses). Superstructure (SS): anisotropic fit of the excess width of $N_{\rm SS}$ superstructure reflections, error = jackknife standard error. $^{*}$Upper error bar set at the largest finite leave-one-out value.}
\label{tab:xi}
\begin{ruledtabular}
\begin{tabular}{cccccc}
$x$ & Host $\xi_{ab}$ & Host $\xi_c$ & SS $\xi_{ab}$ & SS $\xi_c$ & $N_{\rm SS}$ \\ \hline
0.18 & 658 $\pm$ 100 (5) & 1559 $\pm$ 221 (5) & -- & -- & -- \\
0.25 & 606 $\pm$ 107 (5) & 1107 $\pm$ 257 (3) & $313^{+12}_{-11}$ & $207^{+32}_{-24}$ & 9 \\
0.33 & 489 $\pm$ 88 (5) & 1013 $\pm$ 261 (3) & $834^{+1410}_{-322}$ & $119^{+104}_{-38}$ & 8 \\
0.37 & 491 $\pm$ 119 (5) & 1276 $\pm$ 316 (4) & -- & -- & -- \\
0.46 & 644 $\pm$ 91 (6) & 1563 $\pm$ 274 (4) & $416^{+118}_{-75}$ & $110^{+138}_{-39}$ & 7 \\
0.57 & -- & 565 $\pm$ 148 (3) & $728^{+268}_{-154}$ & $1212^{+826\,*}_{-711}$ & 7 \\
\end{tabular}
\end{ruledtabular}
\end{table}

\newpage

\section{Comparison of $c(x)$ with other intercalated dichalcogenides}

\begin{figure}[!h]
    \centering
    \includegraphics[width=1.0\textwidth]{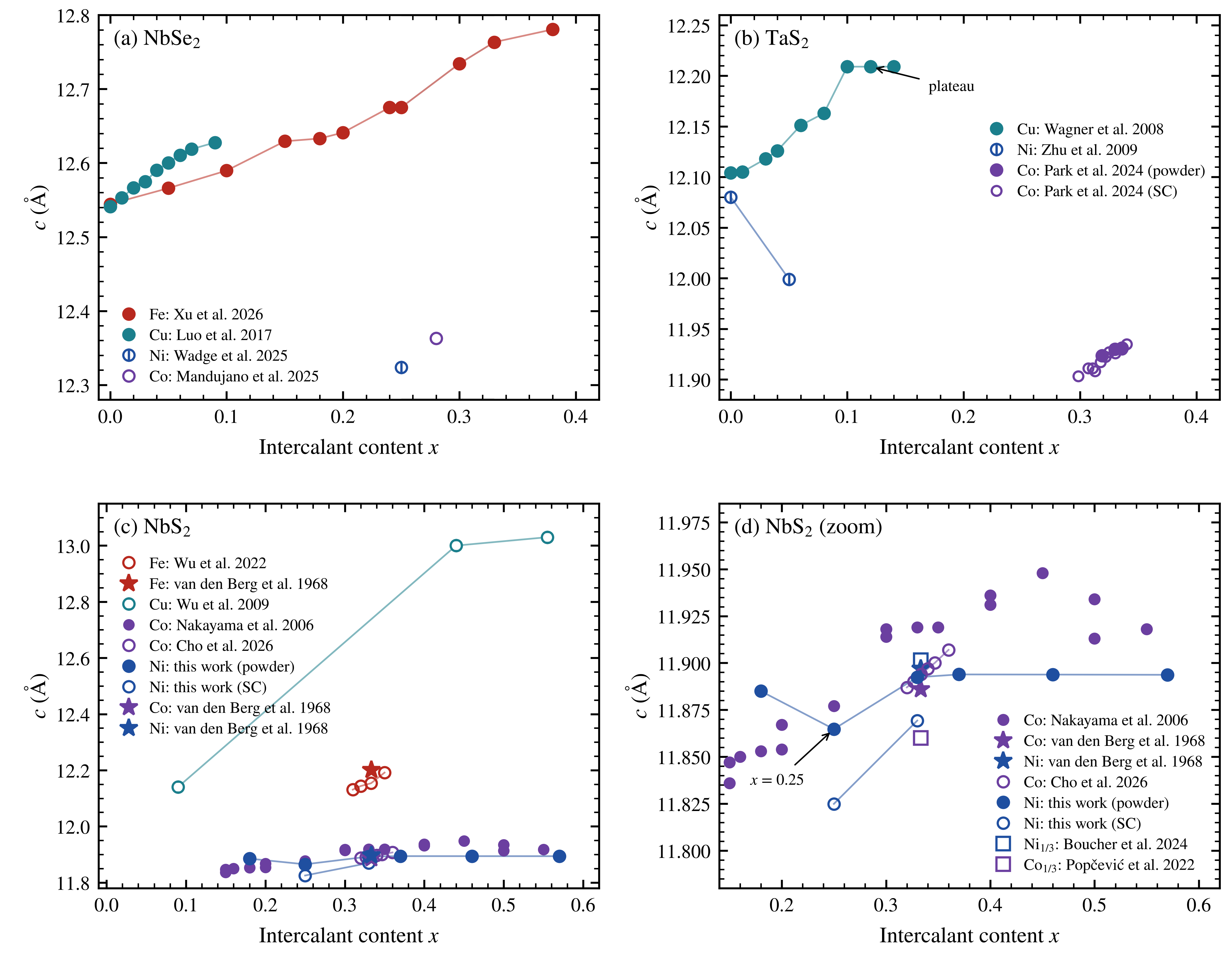}
    \caption{Evolution of the $c$ lattice parameter with intercalant content $x$ in (a) $M_x$NbSe$_2$, (b) $M_x$TaS$_2$ and (c) $M_x$NbS$_2$ ($M$ = Fe, Cu, Co, Ni); panel (d) enlarges the $c\approx11.8$--$11.98$~\AA\ region of (c). Symbol colors denote the intercalant (Fe red, Cu teal, Ni blue, Co purple). Filled symbols are powder diffraction, open symbols single-crystal diffraction, and half-filled symbols data for which the sample type is not stated. Stars mark the $x=1/3$ powder values of van den Berg and Cossee~\cite{vdB68} (as tabulated in Ref.~\cite{Anz70}), and open squares the single-crystal values reported for Ni$_{1/3}$NbS$_2$~\cite{boucher2024} and Co$_{1/3}$NbS$_2$~\cite{popcevic2022}. Our Ni$_x$NbS$_2$ data (blue circles) use the EDX composition and show powder (filled) and single-crystal (open) values. The gray band in (a) marks the $2\times2\to\sqrt3\times\sqrt3$ superstructure crossover. Data are from Ref.~\cite{xu2026} (Fe$_x$NbSe$_2$), \cite{Luo17} (Cu$_x$NbSe$_2$), \cite{Wadge25} (Ni$_x$NbSe$_2$), \cite{Mandujano25} (Co$_{0.28}$NbSe$_2$), \cite{Wagner08} (Cu$_x$TaS$_2$), \cite{Zhu09} (Ni$_x$TaS$_2$), \cite{Park24} (Co$_x$TaS$_2$), \cite{Wu22} (Fe$_x$NbS$_2$), \cite{Wu09} (Cu$_x$NbS$_2$) and \cite{nakayama2006,Cho26} (Co$_x$NbS$_2$). Values from Refs.~\cite{Luo17,Wagner08,Zhu09,Wu09,nakayama2006,Cho26} were digitized from published figures, so that their uncertainty is about the symbol size. Whereas $c$ grows monotonically upon Fe and Cu intercalation, the Ni and Co compounds of NbS$_2$ remain close to the pristine value, and our Ni data show a minimum at $x=0.25$ followed by a plateau for $x\geq0.33$.}
    \label{figphase}
\end{figure}

Figure~\ref{figphase} compares the evolution of the $c$ lattice parameter with intercalant content $x$ in 3$d$-intercalated NbSe$_2$, TaS$_2$ and NbS$_2$.

\section{Structural response to the $2\times2$ order at $x=1/4$}

\begin{table}[!htb]
\centering
\caption{Heights of the S atoms bordering the van der Waals gap above the Ni plane ($h_{\rm Ni}$) and below the adjacent Nb plane ($h_{\rm Nb}$), in \AA, in the refined single-crystal structures and in the relaxed DFT$+U$ supercells. $z$ is the fractional coordinate measured from the Ni plane, so that $h_{\rm Ni}=zc$ and $h_{\rm Nb}=(\tfrac14-z)\,c$. The fraction of the S atoms on each site is given in parentheses.}
\label{tab:Sheights}
\begin{ruledtabular}
\begin{tabular}{llccc}
 & S site & $z$ & $h_{\rm Nb}$ & $h_{\rm Ni}$ \\
\hline
Exp.\ $x=1/4$ & S1 (3/4) & 0.11897(7) & 1.549 & 1.407 \\
              & S2 (1/4) & 0.11436(8) & 1.604 & 1.352 \\
Exp.\ $x=1/3$ & S (all)     & 0.11817(4) & 1.564 & 1.402 \\
DFT\ $x=1/4$  & S1 (3/4) & 0.11912    & 1.551 & 1.412 \\
              & S2 (1/4) & 0.11426    & 1.609 & 1.354 \\
DFT\ $x=1/3$  & S (all)     & 0.11788    & 1.566 & 1.397 \\
\end{tabular}
\end{ruledtabular}
\end{table}

\begin{figure*}[!htb]
\centering
\includegraphics[width=1.0\textwidth]{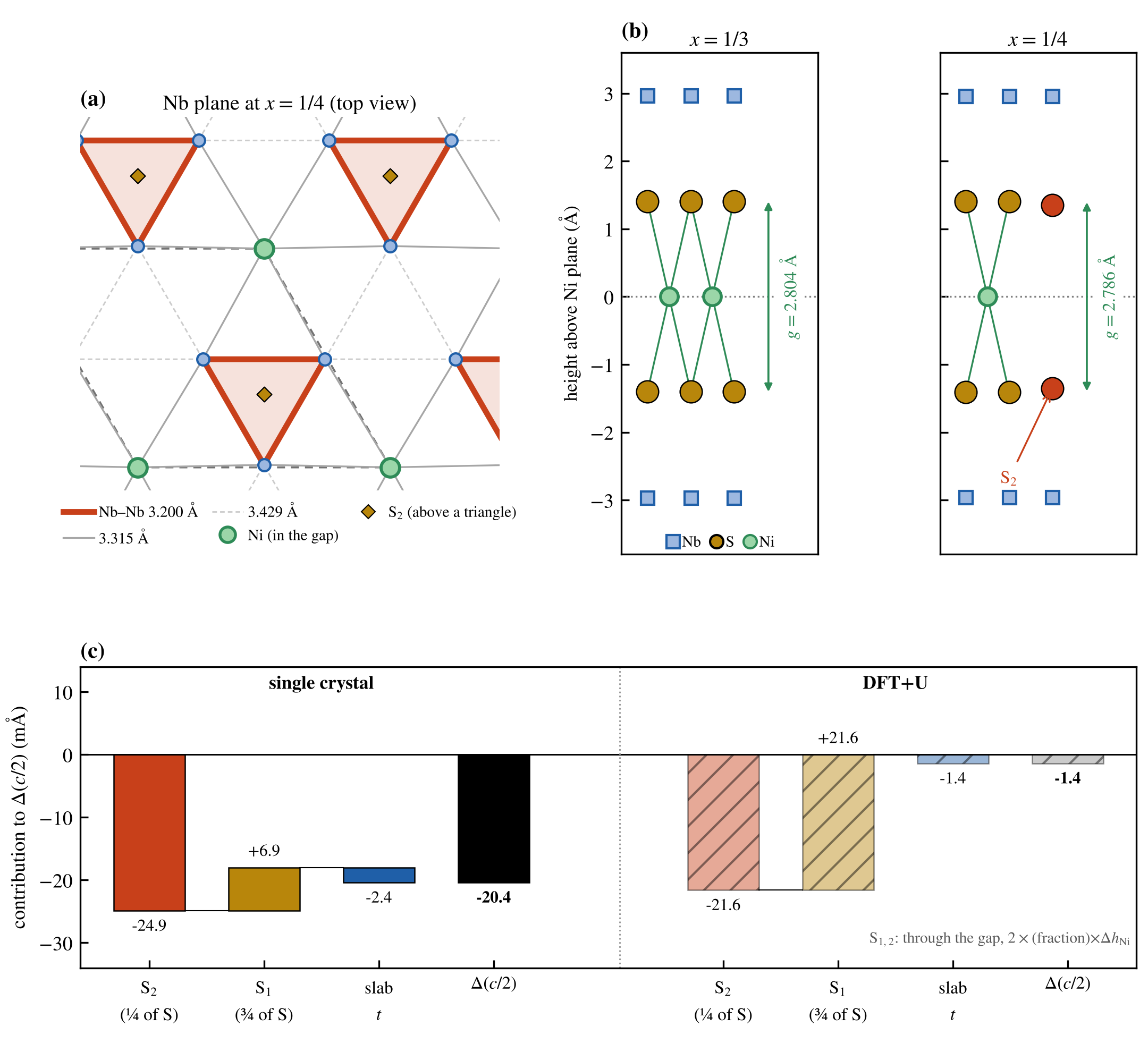}
\caption{Origin of the contraction of $c$ at $x=1/4$ relative to $x=1/3$.
(a)~Nb plane of the $x=1/4$ single-crystal structure, viewed along $c$. The Nb$_3$ triangles contracted to 3.2005(6)~\AA\ (red) are capped by the S2 atoms (diamonds); the other Nb--Nb distances are 3.315 and 3.429~\AA\ (trimerization 3.4\%). Ni atoms (green) sit in the van der Waals gap.
(b)~To-scale side views of the refined structures, with heights measured from the Ni plane. At $x=1/3$ all S atoms are equivalent; at $x=1/4$ they split into S1 (three in four, bonded to Ni) and S2 (one in four, no Ni neighbor), and the lower S2 reduces the mean gap $g$ from 2.804 to 2.786~\AA.
(c)~Contributions to $\Delta(c/2)=c(1/4)/2-c(1/3)/2$, in m\AA: the S2 and S1 height changes, each weighted by twice its fraction of the S atoms, and the S--Nb--S slab thickness $t$.
In the single crystals the S2 shift ($-24.9$) outweighs the outward S1 shift ($+6.9$) and the slab change ($-2.4$), giving $-20.4$~m\AA. In the relaxed DFT$+$U supercells the S1 shift ($+21.6$) cancels that of S2 ($-21.6$), leaving only $-1.4$~m\AA.}
\label{fig:trimer}
\end{figure*}

The $c$-axis dip at $x=1/4$ is accompanied by a distinct distortion of the host. In the $2\times2$ cell the Nb2 atoms ($2b$) lie directly beneath the Ni atoms and remain on the ideal lattice, whereas the Nb1 atoms ($6h$) are displaced by 0.066~\AA\ to form Nb$_3$ trimers with a short bond of 3.2005(6)~\AA\ (Fig.~\ref{fig:trimer}a). Defining the trimerization as $(a'/2-d_{\rm short})/(a'/2)$, where $a'=6.629$~\AA\ is the in-plane lattice parameter of the $2\times2$ cell, so that $a'/2=3.315$~\AA\ is the Nb--Nb distance of the undistorted triangular lattice, we obtain 3.4\,\% at $x=1/4$, compared with $\approx0$ at $x=1/3$. A similar trimerization is found in Fe$_{1/4}$TaS$_2$ (short Ta--Ta bond 3.209~\AA\ against the undistorted $a'/2=3.307$~\AA, i.e.\ 3.0\,\%) \cite{morosan2007}, whose $c$ is also smaller than at $x=1/3$ ($-0.13$~\AA). There the gap shrinks ($g=2.973$ versus 3.071~\AA) while the slab thickens ($t=3.104$ versus 3.071~\AA) \cite{Musfeldt2025}, whereas in Ni$_x$NbS$_2$ the slab changes by only $-2.4$~m\AA.

Defining the interlayer period as $c=2(t+g)$, with $t$ and $g$ averaged over the S sites, the contraction $\Delta(c/2)=-20.4(1)$~m\AA\ ($x=1/4$ relative to $x=1/3$) comes almost entirely from the gap, $\Delta g=-18.0$~m\AA\ ($g=2.786$ versus 2.804~\AA), the slab thickness changing by only $\Delta t=-2.4$~m\AA\ ($t=3.126$ versus 3.128~\AA) (Fig.~\ref{fig:trimer}c). This contraction is not carried by the Ni--S bond, which is slightly longer at $x=1/4$, 2.3841(6) against 2.3796(4)~\AA\ ($\Delta d=+4.5(7)$~m\AA). At $x=1/4$ the S layer bordering the gap splits into S1 ($12k$, bonded to Ni) and S2 ($4f$, bonded to no Ni), which caps the Nb$_3$ trimer. S1 stays within 5~m\AA\ of the height of the S atoms at $x=1/3$, whereas S2 lies 0.050~\AA\ closer to the Ni plane (Table~\ref{tab:Sheights}); the S2 displacement alone contributes $-25$~m\AA\ to $\Delta(c/2)$, partly compensated by $+7$~m\AA\ from the outward shift of S1. The trimerization and the gap contraction thus occur at the same S2 sites. Relaxed DFT$+U$ supercells reproduce both the trimerization and the S2 shift ($-43$~m\AA), supporting a link between the two, but they also move S1 outward by 14~m\AA, which cancels the effect on $g$ and leaves a much weaker minimum of $c$; a direct test of the causal relation would require a bonding analysis of the distorted and undistorted structures.

\section{Examples of Rietveld refinements}

\begin{figure*}[!h]
\centering
\includegraphics[width=1.0\textwidth]{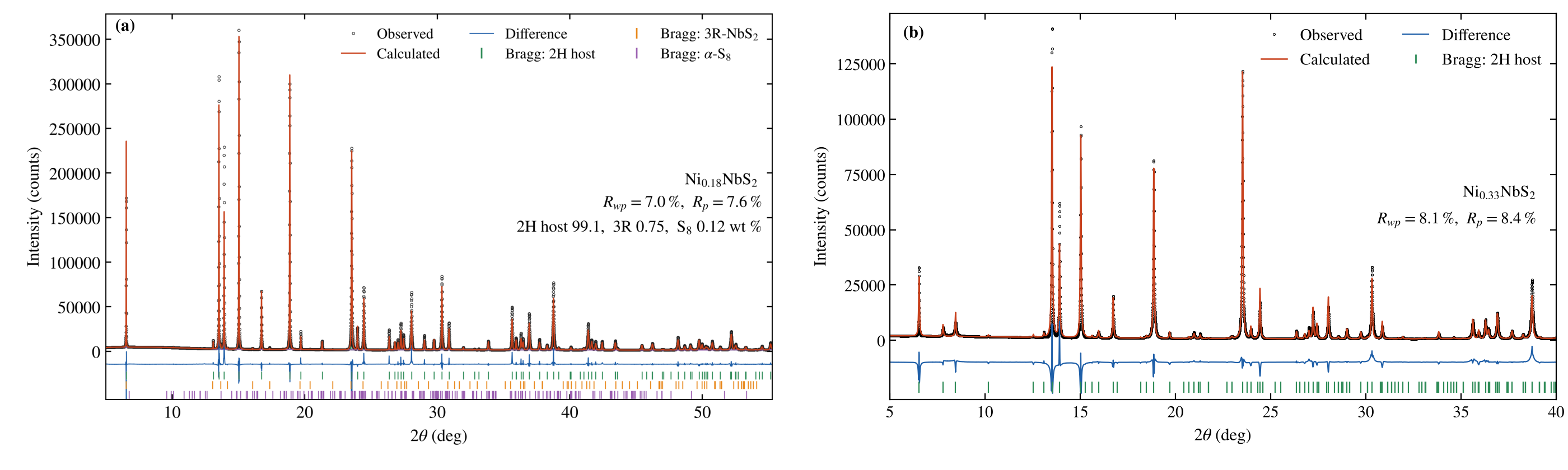}
\caption{Rietveld refinements of the synchrotron powder diffraction patterns ($\lambda=0.6779$~\AA) of (a)~Ni$_{0.18}$NbS$_2$, with the 2H host refined in $P6_3/mmc$ and 3R-NbS$_2$ and $\alpha$-S$_8$ included as secondary phases, and (b)~Ni$_{0.33}$NbS$_2$, with the 2H host refined in the ordered $P6_322$ setting. Open circles: observed; red line: calculated; blue line: difference; vertical ticks: Bragg positions.} 
\label{fig:Refinement}
\end{figure*}

Figure~\ref{fig:Refinement} shows, as examples, the Rietveld refinements of the powder patterns of Ni$_{0.18}$NbS$_2$, in the disordered regime, and of Ni$_{0.33}$NbS$_2$, in the $\sqrt3\times\sqrt3$ ordered regime.

\section{Superlattice intensities and Ni distribution above $x=1/3$}

Table~\ref{tab:SSint} gives the powder intensities of the first even-$\ell$ and odd-$\ell$ reflections of the $\sqrt3\times\sqrt3$ order calculated for several distributions of Ni over the three octahedral sites of the $P6_322$ cell: $2d$, occupied in the ideal structure; $2c$, the sublattice occupied in the neighboring gaps; and $2b$, empty in all gaps of the ideal structure. All calculations use the refined Ni$_{1/3}$NbS$_2$ structure (Table~IV of the main text) for the Nb and S atoms, with $2d$ fully occupied and the excess Ni added on the vacant sites, with a total occupancy $q=3(x-1/3)$ of the $2b$ and $2c$ sites. Ni on $2b$ suppresses the even-$\ell$ reflections and leaves the odd-$\ell$ ones nearly unchanged, Ni on $2c$ does the opposite, and a random distribution over $2b$ and $2c$ preserves their ratio. Table~\ref{tab:SSobs} gives the measured integrated intensities, obtained by fitting each superlattice reflection with a Gaussian on a linear background and each host reflection with a pseudo-Voigt function. Because preferred orientation varies from sample to sample, each superlattice reflection is normalized to a host reflection of similar orientation: $100$ for the $hk0$ reflections and $102$ for $\tfrac13\tfrac13 1$, the scattering vectors of $100$, $102$ and $\tfrac13\tfrac13 1$ making angles of $90^\circ$, $64^\circ$ and $67^\circ$ with $c^*$. The ratio $I(\tfrac13\tfrac13 0)/I(\tfrac13\tfrac13 1)$ is measured within each pattern and requires no normalization. Relative to $x=0.33$, it decreases by a factor of 23 at $x=0.37$ and 33 at $x=0.46$, compared with a factor of 21 calculated for $x=0.46$ with all the excess Ni on $2b$, and no change for a random distribution. The host ratio $I(100)/I(102)$ increases from 1.27 to 1.90 between $x=0.33$ and $0.46$, as calculated for this model (from 0.79 to 1.20), so that preferred orientation affects the two samples alike; its absolute value, about 1.6 times the calculated one in both, gives the order of magnitude of the preferred-orientation correction between reflections at $90^\circ$ and $\approx65^\circ$ from $c^*$. The widths of the $\tfrac13\tfrac13 0$ and $\tfrac13\tfrac13 1$ reflections do not change significantly between $x=0.33$ and $0.46$ (FWHM 0.081 and $0.083^\circ$, and 0.111 and $0.126^\circ$, respectively).

\begin{table}[!htb]
\centering
\caption{Calculated powder intensities ($\lambda=0.6779$~\AA, Cromer--Mann form factors, refined displacement parameters, $I\propto m\,\mathrm{LP}\,|F|^2$, normalized to $I(002)=100$) of the $\tfrac13\tfrac13 0$ and $\tfrac13\tfrac13 1$ reflections, indexed in the parent hexagonal cell, for different occupancies of the $2d$, $2b$ and $2c$ Ni sites of the $\sqrt3\times\sqrt3$ cell.}
\label{tab:SSint}
\begin{ruledtabular}
\begin{tabular}{lccccc}
Model & $n(2d)$ & $n(2b)$ & $n(2c)$ & $I(\tfrac13\tfrac13 0)$ & $I(\tfrac13\tfrac13 1)$ \\ \hline
Ideal, $x=1/3$ & 1 & 0 & 0 & 4.98 & 23.6 \\
Refined crystal, $x=1/3$ & 0.91 & 0.09 & 0 & 2.60 & 19.3 \\
$x=0.37$, excess on $2b$ & 1 & 0.11 & 0 & 3.20 & 25.3 \\
$x=0.37$, excess random & 1 & 0.055 & 0.055 & 4.76 & 22.5 \\
$x=0.46$, excess on $2b$ & 1 & 0.38 & 0 & 0.31 & 30.6 \\
$x=0.46$, excess random & 1 & 0.19 & 0.19 & 4.18 & 19.6 \\
$x=0.46$, excess on $2c$ & 1 & 0 & 0.38 & 12.5 & 11.1 \\
\end{tabular}
\end{ruledtabular}
\end{table}

\begin{table}[!htb]
\centering
\caption{Measured integrated intensities of superlattice reflections (parent hexagonal indexing), normalized to the integrated intensity of a host reflection of similar orientation, $100$ or $102$, and multiplied by $10^3$; ratio of the two $\sqrt3\times\sqrt3$ reflections within each pattern; and host ratio $I(100)/I(102)$ ($\lambda=0.6779$~\AA). Errors are statistical errors of the fits and do not include the systematic uncertainty of the background. n.d.: not detected.}
\label{tab:SSobs}
\begin{ruledtabular}
\begin{tabular}{lcccccc}
$x$ & $\dfrac{I(\tfrac13\tfrac13 0)}{I(100)}$ & $\dfrac{I(\tfrac13\tfrac13 1)}{I(102)}$ & $\dfrac{I(\tfrac13\tfrac13 0)}{I(\tfrac13\tfrac13 1)}$ & $\dfrac{I(\tfrac12 00)}{I(100)}$ & $\dfrac{I(\tfrac12\tfrac12 0)}{I(100)}$ & $\dfrac{I(100)}{I(102)}$ \\[6pt] \hline
0.25 & n.d. & n.d. & -- & 140(3) & 43.7(6) & 1.16 \\
0.33 & 35.8(1.1) & 89(3) & 0.51(2) & n.d. & $<1$ & 1.27 \\
0.37 & 0.84(14) & 54.0(1.6) & 0.022(4) & 16.4(6) & $<1.5$ & 1.43 \\
0.46 & 0.80(8) & 99(3) & 0.015(2) & 7.0(5) & 1.8(1) & 1.90 \\
0.57 & n.d. & $<2$ & -- & 105(4) & 35.8(1.2) & 2.03 \\
\end{tabular}
\end{ruledtabular}
\end{table}

\clearpage

\bibliography{biblio_v8}
\bibliographystyle{apsrev4-1}